\documentclass[12pt,preprint]{aastex}

\shorttitle{Cosmic-Ray Modified Shocks}
\shortauthors{Kang {\it et al.~}}

\def\eg{{\it e.g.,}}
\def\ie{{\it i.e.,~}}

\def\kms{~{\rm km~s^{-1}}}
\def\cm3{~{\rm cm^{-3}}}

\begin{document}
\title{Self-Similar Evolution of Cosmic-Ray Modified Shocks:
The Cosmic-Ray Spectrum}

\author{Hyesung Kang\altaffilmark{1},
        Dongsu Ryu\altaffilmark{2},
    and T. W. Jones\altaffilmark{3}}

\altaffiltext{1}
{Department of Earth Sciences, Pusan National University, Pusan 609-735,
Korea:\\ kang@uju.es.pusan.ac.kr}
\altaffiltext{2}
{Department of Astronomy and Space Science, Chungnam National University,
Daejeon 305-764, Korea:\\ ryu@canopus.cnu.ac.kr}
\altaffiltext{3}
{Department of Astronomy, University of Minnesota, Minneapolis, 
MN 55455, USA:\\ twj@msi.umn.edu}

\begin{abstract}

We use kinetic simulations of diffusive shock acceleration (DSA)
to study the time-dependent evolution of plane,
quasi-parallel, cosmic-ray (CR) modified shocks.
Thermal leakage injection of low energy CRs and finite 
Alfv\'en wave propagation and dissipation are included.
Bohm diffusion as well as the diffusion with the power-law momentum dependence 
are modeled.
As long as the acceleration time scale to relativistic energies is much
shorter than the dynamical evolution time scale of the shocks, the precursor and
subshock transition approach the time-asymptotic state, which depends on 
the shock sonic and Alfv\'enic Mach numbers and the CR injection efficiency.
For the diffusion models we employ, the shock precursor structure evolves in
an approximately self-similar fashion, depending only on the similarity
variable, $x/(u_s t)$. 
During this self-similar stage, the CR distribution at the subshock 
maintains a characteristic form as it evolves: the sum of two power-laws 
with the 
slopes determined by the subshock and total compression ratios
with an exponential cutoff at the highest accelerated momentum, $p_{\rm max}(t)$. 
Based on the results of the DSA simulations spanning a range of Mach numbers,
we suggest functional forms for the shock structure parameters, from which
the aforementioned form of CR spectrum can be constructed.
These analytic forms may represent approximate solutions to the
DSA problem for astrophysical shocks during the self-similar evolutionary
stage as well as during the steady-state stage if $p_{\rm max}$ is fixed.  

\end{abstract}


\section{Introduction}

Diffusive shock acceleration (DSA) is widely accepted 
as the primary mechanism through which cosmic rays (CRs) are
produced in a variety of astrophysical environments
\citep{bell78, dru83, blaeic87}.
The most attractive feature of the DSA theory is the simple prediction of the
power-law momentum distribution of CRs, $f(p) \propto p^{-3\sigma/(\sigma-1)}$ 
(where $\sigma$ is the shock compression ratio) in the
test particle limit. For strong, adiabatic
gas shocks, this gives a power-law index of 4,
which is reasonably close to the observed, `universal' index
of the CR spectra in many environments.

However, it was recognized early on, through both analytical and numerical 
calculations,
that the DSA can be very efficient and that there are highly nonlinear
back-reactions 
from CRs to the underlying flows that modify the spectral form, as well
\citep[\eg][for a review]{maldru01}.
In such CR modified shocks, the pressure from CRs diffusing upstream compresses
and decelerates the gas smoothly before it enters the dissipative subshock,
creating a shock precursor and governing the evolution of the flow velocity 
in the precursor.
On the other hand, it is primarily 
the flow velocity through the precursor and the subshock that controls 
the thermal leakage injection and the DSA of CRs. 
Hence the dynamical structure of the flow and the energy spectrum of CRs
must evolve together, influencing each other in a self-consistent way. 

It is formation of the precursor that causes the momentum distribution of 
CRs to deviate from the simple test-particle power-law distribution.
With a realistic momentum-dependent diffusion, $\kappa(p)$,
the particles of different momenta, $p$, experience different 
compressions, depending on their diffusion length, $l_d(p) =\kappa(p)/u_s$ 
(where $u_s$ is the shock speed).
The particles just above the injection momentum ($p_{\rm inj}$) 
sample mostly the compression across the subshock ($\sigma_s$), 
while those near the highest momentum  ($p_{\rm max}$)
experience the greater, total compression across the entire shock structure
($\sigma_t$).
This leads to the particle distribution function that
behaves as $f(p)\propto p^{-3\sigma_s/(\sigma_s-1)}$ for $p\sim p_{\rm inj}$,
but flattens gradually to $ f(p)\propto p^{-3\sigma_t/(\sigma_t-1)}$ toward
$p\sim p_{\rm max}$ \citep{duffy94}.

Analytic solutions for $f(p)$ at the shock 
have been found in steady-state limits under special conditions;
for example, the case of a constant diffusion coefficient \citep{dru82} and 
the case of steady-state shocks with a fixed $p_{\rm max}$ 
above which particles escape from the system \citep{mal97,mal99,ab05,ab06}.
In these treatments, the self-consistent solutions involve rather
complicated transformations and integral equations, so are difficult to
use in general, although they do provide important insights. 
In particular, \citet{mal99} showed that in highly modified,
strong, steady shocks
($\sigma_t \gg  1$) with a fixed $p_{\rm max}$,
the spectrum of CRs flattens to $f(p)\propto p^{-3.5}$ 
for $\kappa(p) \propto p^{\alpha}$ with $\alpha>1/2$.
He also argued that the form of the CR spectrum is universal under these
conditions, independent of $\kappa(p)$ and $\sigma_t$.  
In an effort to provide more practical description 
\citet{be99} presented a simple approximate model of the CR spectrum at 
strong, steady shocks in plane-parallel geometry. 
They adopted a three-element, piece-wise power-law form to represent 
the spectrum at
non-relativistic, intermediate, and highly relativistic energies.
And they demonstrated that this model approximately represents
the results of their Monte Carlo simulations.

In \citet{kj07} (Paper I), from kinetic equation simulations of 
DSA in plane-parallel shocks with the Bohm-like diffusion ($\kappa \propto p$),
we showed that the CR injection rate and the postshock states approach time-asymptotic
values, even as the highest momentum $p_{\rm max}(t)$ continues to increase with time, 
and that such shocks then evolve in a ``self-similar'' fashion.
We then argued that
the nonlinear evolution of the shock structure and the CR distribution
function in this stage may be described approximately 
in terms of the similarity variables, $\xi = x/(u_s t)$ and 
$Z \equiv {\ln(p/p_{\rm inj}) / \ln[p_{\rm max}(t)/p_{\rm inj}}]$. 
Based on the self-similar evolution, we were able to predict 
the time-asymptotic value of the CR acceleration efficiency 
as a function of shock Mach number for the assumed 
models of the thermal leakage injection and the wave transportation.
In those simulations we assumed that the self-generated waves provide 
scatterings sufficient enough to guarantee the Bohm-like diffusion,
and that the particles do not escape through either an upper momentum 
boundary or a free-escape spatial boundary.
So the CR spectrum extended to ever higher momenta,
but at the same time the particles with the 
highest momentum spread over the increasing diffusion length scale as 
$l_{\rm max}\propto \kappa(p_{\rm max})/u_s \propto p_{\rm max} \propto t$. 
We note that in Paper I we considered plane-parallel shocks with shock
Mach number, $2 \le M_0 \le 80$, propagating into the upstream
gas with either $T_0=10^4$K or  $10^6$ K, since we were interested
mainly in cosmic structure formation shocks.

The simplicity of the results in Paper I suggested that it might be 
possible to obtain an approximate analytic expression for the CR spectrum 
in such shocks, but the simulations
presented in that paper were not sufficient to address that question.
Thus we further carried out an extensive set of simulations
to explore fully the time-dependent behavior of the CR distribution in
CR modified shocks with shock Mach numbers $M_0\ge 10$.
In this paper, from the results of these simulations,
we suggest practical analytic expressions 
that can describe the shock structure and the energy spectrum of accelerated particles
at evolving CR modified shocks in plane-parallel geometry,
in which the Bohm-like diffusion is valid.  

In realistic shocks, however, once the diffusion length $l_{\rm max}$ becomes comparable 
to the curvature of shocks, or when the growth of waves generated by the CR streaming
instability is inefficient, 
the highest energy particles start to escape from the system before they are scattered and 
advected back through the subshock. 
In such cases, $p_{\rm max}$ is fixed, and the CR spectrum and the shock structure 
evolve into steady states.
So, for comparison, we carried out additional simulations for analogous shocks 
in which the particles are allowed to escape from the system once they are accelerated
above an upper momentum boundary, $p_{\rm ub}$.
Those shocks achieve true steady states and  
the shock structure and the CR distribution become stationary with forms similar 
to those maintained during the self-similar stage of shock evolution.
In this sense, our solution is consistent with the analytic solutions
for steady state shocks obtained in the previous papers mentioned above.

In the next section we describe the numerical simulations and results. 
The approximate formula for the CR spectrum will be  presented and
discussed in \S 3, followed by a summary in \S 4. We also include an
appendix that presents simple analytic and empirical expressions that
can be used to characterize the dynamical properties of CR modified shocks. 


\section{Numerical Calculations}

\subsection{Basic equations}

In our kinetic simulations of DSA,
we solve the standard gasdynamic equations with the CR pressure terms
in the conservative, Eulerian form
for one-dimensional plane-parallel geometry
\citep{kjg02,kj05,kj07},
\begin{equation}
{\partial \rho \over \partial t}  +  {\partial (u \rho) \over \partial x} = 0,
\label{masscon}
\end{equation}
\begin{equation}
{\partial (\rho u) \over \partial t}  +  {\partial (\rho u^2 + P_g + P_c) \over
\partial x} = 0,
\label{mocon}
\end{equation}
\begin{equation}
{\partial (\rho e_g) \over \partial t} + {\partial \over \partial x}
(\rho e_g u + P_g u) =
-u {{\partial P_c}\over {\partial x}} + W(x,t) - L(x,t),
\label{econ}
\end{equation}
where $P_g$ and $P_c$ are the gas and CR pressures,
respectively, $e_g = {P_g}/{[\rho(\gamma_g-1)]}+ u^2/2$
is the total gas energy per unit mass.
The remaining variables, except for $L$ and $W$, have the usual meanings.
The injection energy loss term, $L(x,t)$, accounts for the
energy carried away by the suprathermal particles injected into the CR
component at the subshock and is subtracted from the postshock gas
immediately behind the subshock.
The gas heating due to the Alfv\'en wave dissipation in the upstream region
is represented by the term 
\begin{equation}
W(x,t)= - v_A \frac{\partial P_c}{ \partial x },
\label{wheat}
\end{equation}
where $v_A= B/\sqrt{4\pi \rho}$ is the local Alfv\'en speed (Paper I).
These equations can be used to describe parallel shocks, where the
large-scale magnetic field is aligned with the shock normal and
the pressure contribution from the turbulent magnetic fields
can be neglected.

The CR population is evolved by solving the diffusion-convection equation 
for the pitch-angle-averaged distribution function, $f(x,p,t)$,
in the form,
\begin{equation}
{\partial g\over \partial t}  + (u+u_w) {\partial g \over \partial x}
= {1\over{3}} {\partial \over \partial x} (u+u_w) \left( {\partial g\over
\partial y} -4g \right) + {\partial \over \partial x} \left[\kappa(x,y)
{\partial g \over \partial x}\right],
\label{diffcon}
\end{equation}
where $g=p^4f$ and $y=\ln(p)$ \citep{ski75a}.
Here, $\kappa(x,p)$ is the spatial diffusion coefficient.
The CR population is isotropized with respect to the local Alfv\'enic
wave turbulence, which would in general move at a speed $u_w$ with respect
to the plasma.
Since the Alfv\'en waves upstream of the subshock are expected to be
established
by the streaming instability, the wave speed is set there to be $u_w=v_A$.
Downstream, it is likely that the Alfv\'enic turbulence is nearly
isotropic, so we use $u_w=0$ there.

We consider two models for CR diffusion: Bohm diffusion and
power-law diffusion, 
\begin{eqnarray}
\kappa_B = \kappa^* \left(\frac{\rho_0}{\rho}\right)^{\nu}
\frac{p^2}{\sqrt{p^2+1}},
\nonumber \\ 
\kappa_{pl} = \kappa^* \left(\frac{\rho_0}{\rho}\right)^{\nu} p^{\alpha}, 
\label{kappa}
\end{eqnarray}
with $\alpha=0.5-1$. 
Hereafter, the momentum is expressed in units of $m_pc$, where $m_p$
is the proton mass and $c$ is the speed of light. So,  $\kappa^*$ is
a constant of dimensions of length squared over time.
As in our previous studies,
we consider diffusion both without and with 
a density dependence, $\rho_0/\rho$; that is, either $\nu = 0$ or
$\nu = 1$. The latter case quenches the CR acoustic instability \citep{dru84}
and approximately accounts for the compressive amplification of Alfv\'en waves.
Since we do not follow explicitly the amplification of magnetic fields due to
streaming CRs,
we simply assume that the field strength scales with compression and
so the diffusion coefficient scales inversely with density.
Bohm-like diffusion is an idealization of what is expected in a dynamically
evolving CR modified shock. As discussed in \S 2.3 the diffusion coefficient, which
results from resonant scattering with Alfv\'en waves, varies inversely with the intensity
of the resonant waves. The wave intensity is expected to be amplified as the shock evolves
from upstream, ambient values via the streaming instability. Bohm diffusion represents
the simplest nonlinear limited model for that process. The very highest momentum
CRs will encounter ambient wave intensities, so perhaps below levels implied by
Bohm diffusion. The model assumes that the streaming instability quickly
amplifies those waves to nonlinear levels \citep[\eg][]{ski75b, lucek00}.
We label the quantities upstream of the
shock precursor by the subscript `0', those immediately
upstream of the gas subshock by `1', and those downstream by `2'.
So, $\rho_0$, for example, stands for the density of the upstream gas.

Equations (1), (2), (3), and (5) are simultaneously integrated by the CRASH
(Cosmic-Ray Acceleration SHock) code. 
The detailed description of the CRASH code can be found in \citet{kjg02} and Paper I.
Three features of CRASH are important to our discussion below. First,
CRASH applies an adaptive mesh refinement technique around the subshock.
So the precursor structure is adequately resolved to couple
the gas to the CRs of low momenta, whose diffusion lengths can be
at least several orders of magnitude smaller that the precursor width. 
Second, CRASH uses a subgrid shock
tracking; that is, the subshock position is followed accurately within a single
cell on the finest mesh refinement level. Consequently, the effective
numerical subshock thickness needed to compute the spatial derivatives
in equation (\ref{diffcon}) is always less than the single cell size of the
finest grid. Third, we calculate the exact subshock speed
at each time step to adjust the rest frame of the
simulation, so that the subshock is kept inside the same grid cell
throughout. These three features enable us to obtain good numerical
convergence in our solutions with a minimum of computational efforts.
As shown in Paper I, the CRASH code can obtain reasonably converged
dynamical solutions even when the grid spacing in the finest refined level 
is greater than the diffusion length of the lowest energy particles 
(\ie $\Delta x_8 > l_d(p_{\rm inj}$)).
This feature allows us to follow the particle acceleration for a large
dynamic rage of $p_{\rm max}/p_{\rm inj}$, typically, $\sim 10^9$,
although the evolution of the energy spectrum at low energies 
and the early dynamical evolution of the shock structure may not be
calculated accurately.

\subsection{Simulation Set-up}

The injection and acceleration of CRs at shocks depend in general
upon various shock parameters such as the Mach number,
the magnetic field strength and obliquity angle, and the
strength of the Alfv\'en turbulence responsible for scattering.
In this study we focus on the relatively simple case of
CR proton acceleration at quasi-parallel shocks, which is appropriately
described by equations (\ref{masscon}) - (\ref{econ}).
The details of simulation set-up can be found in Paper I,
and only a few essential features are briefly summarized here.
Except for diffusion details, the set-up
described here is identical to those reported in Paper I.

As in Paper I, a shock is specified by the upstream gas temperature 
$T_0 $ and the initial Mach number $M_0$.
Two values of $T_0$, $10^4$ K and $10^6$ K, are considered,
representing the warm photoionized gas and
the hot shock-heated gas often found 
in astrophysical environments, respectively.
Then the initial shock speed is given as
\begin{equation}
u_{s,i} = c_{s,0} M_0 = 15\kms \left({T_0 \over 10^4} \right)^{1/2} M_0,
\end{equation}
where $c_{s,0}$ is the sound speed of the upstream gas.
All the simulations reported in this paper have $M_0 = 10$, which is
large enough to produce significant CR modification.
In Paper I we considered a wide range of shock Mach numbers and 
examined the Mach-number dependence of the evolution of CR modified shocks. 
The CR injection and acceleration efficiencies are determined mainly
by the sonic Mach number and the relative Alfv\'en Mach number for 
shocks with $M_0 \ga 10$ \citep{kjg02, kang03}.
On the other hand, they depend sensitively on other model parameters for 
shocks with lower Mach numbers. 
In this paper we thus focus on the evolution of the CR spectrum at moderately strong 
shocks with $M_0 \ga 10$.  We will consider the more complicated problem of weaker shocks in a 
separate paper. 

In our problem, three normalization units are required for length,
time, and mass.
While ordinary, one-dimensional, ideal gasdynamic problems do not contain any intrinsic 
scales, the diffusion in the DSA problem introduces one;
that is, either a diffusion length or a diffusion time,
which of course depend on the particle momentum.
So let $p^{\dag}$ be a specific value of the highest momentum that
we aim to achieve by the termination time of our simulations.
Then the greatest width of the precursor
is set by the diffusion length of the particles with $p^{\dag}$,
$l_d(p^{\dag}) =\kappa(\rho_0,p^{\dag})/u_s$,
while the time required
for the precursor to reach that width is given by 
$t_{\rm acc} (p^{\dag}) \propto l_d(p^{\dag})/u_{s}$ 
(see eq. [\ref{tacc}]).
Hence we choose diffusion length and time for $p^{\dag}$, 
$\hat x = \hat\kappa/\hat u$ and $\hat t = \hat\kappa/\hat u^2$,
with $\hat u = u_{s,i}$ and $\hat\kappa =\kappa(\rho_0,p^{\dag})$,
as the normalization units for length and time.
For the normalization units for mass, we choose $\hat \rho = \rho_0$.
Then the normalized quantities become $\tilde x = x/\hat x$, 
$\tilde t = t/\hat t$, $\tilde u = u/\hat u$,
$\tilde \kappa = \kappa /\hat\kappa$, and $\tilde \rho = \rho/\hat\rho$.
In addition, 
the normalized pressure is expressed as $\tilde P = P/(\hat \rho \hat u^2)$.
With these choices, we expect that at time $\tilde t \sim 1$, the
precursor width would be $\tilde x \sim \tilde l_d(p^{\dag}) \sim 1$, for example.
It should be clear that the physical contents of our normalization are 
ultimately determined by the value of $p^{\dag}$ anticipated
to correspond to $\tilde t \sim 1$ as well as by the form of $\kappa(\rho,p)$.
In the simulations reported here, $p^{\dag}$ was selected
to give us the maximum span of $p$ that is consistent with our
ability to obtain converged results with available computational
resources. Our choice of $p^{\dag}$
is especially dependent on the nonrelativistic momentum dependence of
$\kappa(p)$. In particular, when the dependence is steep, 
$\kappa(p_{\rm inj})$ and $l_d(p_{\rm inj})$ can become extremely small compared to
their relativistic values, necessitating very fine spatial resolution
around the subshock.

In Table 1, we list our numerical models classified by 
$T_0$ and $\kappa$. For example, T6P1 model adopts
$T_0=10^6$ K and $\kappa_{pl}$ with $\alpha=1$  and $\nu = 0$,
while T4Bd model adopts $T=10^4$ K and Bohm diffusion, $\kappa_B$, 
with $\nu = 1$.
In the power law diffusion models of T6P1 and 
T6P1d, $p^{\dag} \sim 10^6$ is chosen for the normalization, 
so that $\tilde \kappa(\tilde\rho=1) = \tilde\kappa^* p = 10^{-6}p$. 
For the Bohm diffusion models, T6Bd and T4Bd, on 
the other hand, $p^{\dag} \sim 10^2$ is chosen, because the steep 
nonrelativistic form of the diffusion makes those models too
costly for us to follow evolution to much higher CR momenta.

A specific example can clarify the application of these simulations to real
situations. 
Let us consider a shock with $u_{s,i}=1.5\times10^3~{\rm km ~s^{-1}}$ 
propagating into the interstellar medium with $B=5~\mu$G.
Then in the Bohm limit that the relativistic CR scattering length equals the
gyroradius, $\kappa^* = m_p c^2/(3eB) = 6.3 \times 10^{21}~{\rm cm^2~s^{-1}}$.
For the T6P1 model, for instance, the
normalization constants are $\hat u= 1.5\times 10^3~ {\rm km ~s^{-1}}$ and
$\hat \kappa= 6.3\times10^{27}~ {\rm cm^2~s^{-1}}$, so
$\hat x= 4.2\times10^{19}$ cm and $\hat t=2.8\times 10^{11}\ {\rm s}$.

On the other hand, the time evolution of these shocks becomes approximately
self-similar, as we will demonstrate. In that case the normalization choices
above are entirely for the convenience of computation. 
We will eventually replace even these normalized physical
variables with dimensionless similarity variables.  
To simplify the notation in the meantime, we hereafter drop the
{\it tilde} from the normalized quantities as defined above.

Our simulations start with a purely gasdynamic shock of $M_0=10$ at rest
at $x = 0$, initialized according to 
Rankine-Hugoniot relations with $u_0 = -1$, $\rho_0 = 1$ and
a gas adiabatic index, $\gamma_g = 5/3$.
So the initial shock speed is $u_{s,i}=1$ in code units.
There are no pre-existing CRs, \ie $P_c(x)=0$ at $t=0$.

\subsection{Thermal leakage and Alfv\'en wave transport}

Although the shock Mach number is the key parameter that determines
the evolution of CR modified shocks, the thermal leakage injection
and the Alfv\'en wave transport are important elements of DSA. 
They were discussed in detail in previous papers including Paper I. 
So here we briefly describe only the central concepts
to make this paper self-contained. 

In the CRASH code, the injection of suprathermal particles via thermal leakage 
is emulated numerically by adopting a ``transparency function'', 
$\tau_{\rm esc}(\epsilon_B, \upsilon)$,
which expresses the probability of downstream particles at given random velocity, 
$\upsilon$, successfully swimming upstream across the subshock through the
postshock MHD waves \citep{kjg02}, whose amplitude is parameterized by $\epsilon_B$.
Once such particles cross into the upstream flow, they are subject to scattering
by the upstream Alfv\'en wave field, so participate in DSA.
The condition that 
non-zero probability for suprathermal downstream particles to 
cross the subshock (\ie $\tau_{\rm esc}>0$ for $p>p_{\rm inj}$)
effectively selects the lowest momentum of the
particles entering the CR population.
The velocity $\upsilon$ obviously must exceed the flow speed of 
the downstream
plasma, $u_2$. In addition, leaking particles must swim against the
effective pondermotive force of MHD turbulence in the downstream plasma. 
The parameter, $\epsilon_B = B_0/B_{\perp}$ used to represent this,
is the ratio of the magnitude of the large-scale magnetic field
aligned with the shock normal, $B_0$, to the amplitude of the postshock 
wave field that interacts with low energy particles, $B_{\perp}$.
It is more difficult for particles to swim upstream
when the wave turbulence is strong ($\epsilon_B$ is small), leading
to smaller injection rates.
\citet{mv98} argued on plasma physics grounds that it should be
$0.25\lesssim\epsilon_B\lesssim 0.35$.
Our own CR shock simulations established that 
$\epsilon_B\sim 0.2-0.25$ leads to 
injection fractions in the range of $\sim 10^{-4}-10^{-3}$,
which are similar to the commonly adopted values in other models 
\citep[\eg][]{mal97,ab05}.
In this study, we use $\epsilon_B=0.2$ for numerical models, although
the choice is not critical to our conclusions.

The CR transport in DSA is controlled by the intensity, spectrum and
isotropy of the Alfv\'enic turbulence resonant with CRs. 
Upstream of the subshock, the Alfv\'enic turbulence
is thought to be excited by the streaming CRs \citep[\eg][]{bell78,lucek00}.
Recently there has been much emphasis on the possible amplification of
the large-scale magnetic field via non-resonant wave-particle interactions
within the shock precursor \citep[\eg][]{bell04,ab06,vlad06}.
Those details will not concern us here; we make the simplifying
assumption that the Alfv\'enic turbulence saturates and that
scattering isotropizes the CR distribution in the frame moving
with the mean Alfv\'en wave motion (see eq. [\ref{diffcon}]). 
Since the upstream
waves are amplified by the CRs escaping upstream, the wave
frame propagates in the upstream direction; \ie $u_w > 0$. Downstream,
various processes should isotropize the Alfv\'en waves 
\citep[\eg][]{ach86}, so the
wave frame and the bulk flow frame coincide; \ie $u_w = 0$. 
This transition in $u_w$ across the subshock
reduces the velocity jump experienced by CRs during DSA.
Since it is really the velocity jump rather than the density jump
that sets the momentum boost, this reduces the acceleration rate
somewhat when the ratio of the upstream sound speed to the Alfv\'en
speed is finite.
An additional effect that has important impact is
dissipation of Alfv\'en turbulence stimulated by the
streaming CRs. That energy heats the inflowing plasma beyond
adiabatic compression.
The detailed physics is complicated and nonlinear, but
we adopt the common, simple assumption that the dissipation
is local and that the wave growth saturates, so that the
dissipation rate matches the rate of wave stimulation
(see eq. [\ref{wheat}]) \citep{jon93, berz97}.
This energy deposition increases the sound speed of the precursor gas, 
thus reducing the Mach number of the flow into the subshock, 
again weakening DSA to some degree \citep[\eg][]{ach82}.
Thus, the CR acceleration becomes less efficient, when the Alfv\'en wave drift
and heating terms are included \citep{berz97,kj06}. 

The significance of these effects can be parameterized by
the ratio of the magnetic
field to thermal energy densities, $\theta =  E_{B,0}/E_{th,0}$, in the
upstream region, which scales as the square of the ratio of the
upstream Alfv\'en ($\upsilon_A$) and sound speeds.
In Paper I, we considered $0.1\le \theta \le 1$; 
here we set $\theta=0.1$.
The dependence of shock behaviors on that parameter are
outlined in Paper I.
The $\theta$ parameter can be related to the more commonly
used shock Alfv\'enic Mach number, $M_{A,0} = u_{s,i}/v_{A,0}$, and
the initial sonic Mach number, $M_0$, as
$M_{A,0} = M_0 \sqrt{\gamma_g(\gamma_g - 1)/(2\theta)}$, where
$v_{A,0} = B_0/\sqrt{4\pi \rho_0}$. With $\gamma_g = 5/3$ and $\theta = 0.1$,
this translates into $M_{A,0} = 2.36 M_0$. So, for our $M_0 = 10$
shocks, $M_{A,0} \approx 24$. Our initial shock speeds are 
$u_{s,i} = 150~{\rm km~s}^{-1}$ for $T_0 = 10^4$ K and  
$u_{s,i} = 1500~{\rm km~s}^{-1}$ for $T_0 = 10^6$ K, corresponding,
then, to $v_A = 6.4~{\rm km~s}^{-1}$ and  $v_A = 64~{\rm km~s}^{-1}$,
respectively. For our example magnetic field, $B_0 = 5\ \mu G$, the
associated upstream gas density would be 
$\rho_0 \approx 5\times 10^{-24}~{\rm g~cm}^{-3}$ and 
$\rho_0 \approx 5\times 10^{-26}~{\rm g~cm}^{-3}$, respectively. 

\section{Results}

\subsection{Evolution toward an asymptotic state}

In the early evolutionary stage,
as CRs are first injected and accelerated at the subshock,  
upstream diffusion creates a CR pressure gradient that
decelerates and compresses the inflowing gas within
a shock precursor. This leads to a gradual decrease of
shock speed with respect to the upstream gas (Fig. 1 [a]-[b]).
As the subshock consequently weakens, the CR injection rate decreases 
due to a reduced velocity jump across the subshock. 
The CR spectrum near $p_{\rm inj}$ also steepens (Fig. 1 [c]-[d]).
The total compression across the entire shock structure actually
increases 
to about 5 in the Mach 10 shocks reported here. The
highest momentum CRs respond to the total shock transition, which
flattens the spectrum at higher momenta; \ie the CR spectrum evolves 
the well-known concave curvature between the lowest and the highest 
momenta. 
Each of these evolutionary features continue to be enhanced 
until preshock compression, CR injection 
at the subshock, and CR acceleration through the entire shock structure all reach 
{\it self-consistent dynamical equilibrium states}  (Fig. 1 [e]-[f]).
Once compression in the precursor reaches the level at which DSA begins to saturate, 
meaning the reduced subshock strength reduces CR injection to maintain 
an equilibrium, the shock compression
($\sigma_s=\rho_2/\rho_1$ and $\sigma_t = \rho_2/\rho_0$)
as well as the gas and CR pressures should remain approximately constant 
during subsequent shock evolution. 
From that time on the structure of the precursor
and the CR spectrum must evolve in tandem to maintain these dynamical features.

The CR pressure is calculated from the particle distribution function by
\begin{equation}
P_c  = { {4\pi}\over 3} m_p c^2 \int_{p_{\rm inj}}^{\infty} g(p)
{p \over \sqrt{p^2+1}}
{dp \over p}.
\label{pcint}
\end{equation}
To see how $P_c$ evolves during the early, nonrelativistic stage,
consider the idealized the test-particle case where the CR 
distribution has a power-law form,
$g(p)= g_0(p/p_{\rm inj})^{-\delta}$ up to $p=p_{\rm max}$,
where $0< \delta \equiv (4-\sigma_s)/(\sigma_s-1)<0.5$ for 
the shock compression ratio of $4>\sigma_s>3$. 
Then one can roughly express 
$P_c \propto [(p_{\rm max}/p_{\rm inj})^{1-\delta} -1]
\propto (p_{\rm max}/p_{\rm inj})^{1-\delta} $ for $p_{\rm inj} 
\ll p_{\rm max}<1$.
In a strong, unmodified shock, $1-\delta \approx 1$, and $P_c$ 
initially increases quickly as
$P_c \propto p_{\rm max}/p_{\rm inj}$.
We will show in \S 3.3, as the shock becomes modified toward 
the dynamical equilibrium state, that
the CR pressure is dominated by relativistic particles and 
the CR spectrum evolves in a manner that leads to nearly 
constant postshock $P_{c,2}$. 
These features in the evolution of $P_{c,2}$ are illustrated 
in Figure 1 (e) - (f).
The time-asymptotic states are slightly different among different models,
because the numerically realized CR injection rate depends weakly 
on $\kappa(p)$.

The mean acceleration time for a particle to reach $p_{\rm max}$ from $p_{\rm inj}$ 
in the test-particle limit of DSA theory is given by \cite[\eg][]{dru83}
\begin{equation}
t_{\rm acc} = {3\over {u_0-u_2}} \int_{p_{\rm inj}}^{p_{\rm max}}
\left({\kappa_0\over u_0} + {\kappa_2\over u_2} \right) {dp \over p}.
\label{tacc}
\end{equation}
For power-law diffusion with density dependence,
$\kappa_{pl}=\kappa_* p^{\alpha} (\rho_0/\rho)^{\nu}$,
the maximum momentum can be estimated by setting $t = t_{\rm acc}$ as
\begin{equation} 
p_{\rm max}(t) \approx \left[ {{\alpha (\sigma_t -1)} 
\over {3\sigma_t(1+\sigma_t^{1-\nu})}}
{u_s^2 \over \kappa^*} t  \right]^{1/\alpha} 
= \left[ f_c {u_s^2 \over \kappa^*} t  \right]^{1/\alpha},
\label{p1a}
\end{equation}
where $f_c \equiv \alpha (\sigma_t -1) / 
\left[3\sigma_t(1+\sigma_t^{1-\nu}) \right]$ is a  
constant factor during the self-similar stage 
and $u_s$ is the shock speed in the time-asymptotic limit.
As the feedback from CRs becomes important, 
the shock speed relative to far upstream flow is reduced, 
typically about 10-20 \% for the shock parameters considered here 
(\ie $u_s\approx [0.8-0.9] u_{s,i}$). 
With $\alpha=1$ and $\nu=1$, for a typical value of $\sigma_t\approx 5.3$ 
for a $M_0=10$ shock, $f_c \approx 0.13$.

In an evolving CR shock, at a given shock age of $t$, 
the power-law spectrum should extend 
roughly to $p_{\rm max}(t)$ above which it should decrease exponentially. 
Then the diffusion length of the most energetic particles 
increases linearly with time as 
\begin{equation} 
l_{\rm max} (t) \equiv {{\kappa^* p_{\rm max}^{\alpha}(t) } \over u_s} = 
f_c u_s t.
\label{lmax}
\end{equation}
So $l_{\rm max}(t)$ depends only on the characteristic length $u_s t$, 
independent of the size of the diffusion coefficient,
although at a given time the particles are accelerated to higher energies 
with smaller values of $\kappa^*$. 
Since the precursor scale height is proportional to $l_{\max}$, 
the precursor broadens linearly with time,
again independent of the size of $\kappa^*$.
This is valid even for the Bohm diffusion if $p_{\rm max}\gg1$, 
since $\kappa_B \approx \kappa^* p$ for $p\gg1$.
Thus, the hydrodynamic structure of evolving CR shocks does not depend
on the diffusion coefficient,
even though the CR diffusion introduces the diffusion length and time
scales in the problem.

\subsection{Shock structure and CR spectrum in self-similar stage}

After the precursor growth reaches a time-asymptotic form,
the shock structure follows roughly the self-similar evolution and
stretches linearly with time, as noted above. 
Thus, we show in Figure 2 the evolution of a $M_0=10$ shock with T6P1d model 
in terms of the similarity variable, $\xi = x/(u_{s,i}t)$, 
for $t>1$ (\ie later stage of the shock shown in Fig. 1). 
The time-asymptotic shock speed approaches $u_s = u_0  \approx 0.9 u_{s,i}$
for these shock parameters. The reduction in shock speed results from
the increase in $\sigma_t$, so depends upon the degree of shock modification.
Here $\sigma_t \approx 5.3$, $\alpha = 1$, $\nu = 1$, 
so equation (\ref{lmax}) give
$l_{\rm max} \approx 0.13 u_s t$, which corresponds to
the precursor scale height in terms of $\xi$, $H_{\xi} \equiv l_{\rm max}/(u_{s,i} ~t) \approx 0.12$.

We also show the approximate self-similar evolution of the shock structure 
for four additional models with $\kappa(\rho, p)$ listed in Table 1 (Fig. 3). 
As discussed in \S3.1, the overall shock structure at a given time $t$
is roughly independent of the diffusion coefficient,
except for some minor details in the shock profile that have developed in
the early stage.
Also the shock evolution seems to be approximately self-similar 
in all the models, as shown in the middle and right panels of Figure 3. 
Of course, with different values of $\kappa^*$ and $\alpha$,
on the other hand, the highest momentum of the CR spectrum at a given
time depends on $\kappa$ (see Fig. 5).

Figure 4 (a)-(b) shows how the particle distribution at the subshock,
$g_s(p)=f(x_s, p) p^4$, evolves during the self-similar stage, extending 
to higher $p_{\rm max}$.
For this model equation (\ref{p1a}) gives
$p_{\rm max} \approx (0.1/\kappa^*)~ t = 10^5 ~t$.
This estimate is quite consistent with the evolution of $g_s(p)$ shown
in this figure.
The peak value of $g_s(p)$ near $p_{\rm max}$ seems to 
remain constant during the self-similar stage. 
This reflects the fact that $P_{c,2}$ remains
constant, as it must once DSA is saturated, and the fact that $P_c$ is
dominated by relativistic CRs near $p_{\rm max}$ for strong shocks.

The injection momentum, $p_{\rm inj} \propto \sqrt{P_{g,2}/\rho_2}$,
becomes constant in time after the initial adjustment,
because the postshock state is fixed in the self-similar evolution stage. 
Then the value of $g_s(p_{\rm inj})$ is fixed by 
$g_{s,th}(p_{\rm inj})$, the thermal distribution of the postshock gas 
at $p_{\rm inj}$, and stays constant, too. 

Let us suppose particles with a given momentum $p_1$ experience on average 
the velocity jump over the diffusion length $\xi_1=l_d(p_1)/(u_s ~t_1)$, 
$\Delta u(\xi_1)$, at time $t_1$.   
At a later time $t$ they will be accelerated to $p= p_1 \cdot (t/t_1)^{1/\alpha}$
and diffuse over the scale, $\xi=l_d(p)/(u_s ~t) = \xi_1$.
So they experience the same velocity jump $\Delta u(\xi_1)$,
as long as the velocity profile, $u(\xi)$, remains constant during the self-similar stage. 
Then the spectral slopes plotted in terms of $p/p_{\rm max}$ should 
retain a similar shape over time.
The slope of the distribution function at the subshock, $q= - d \ln g_s/ d \ln p+4$, 
and the slope of the volume integrated distribution function,
$Q= - d \ln G/ d \ln p + 4$ (where $G=\int g dx$), 
as a function of $p/p_{\rm max}(t)$ are shown in Figure 4 (d).
Low energy particles near $p_{\rm inj}$ experience the subshock compression
only, while highest momentum particles near $p_{\rm max}$ feel the total
shock compression. 
So $q(p) \approx q_s= 3\sigma_s/(\sigma_s-1)$ for $p\sim p_{\rm inj}$,
while $q(p) \approx q_t =3\sigma_t/(\sigma_t-1)$ for $p\sim p_{\rm max}$. 
The numerical results are roughly consistent with such expectations.

Consequently, to a good approximation, $g_s(p)$ evolves with fixed amplitudes, 
$g_s(p_{\rm inj})$ and $g_s(p_{\rm max}$), and with fixed spectral slopes,
$q_s$ and $q_t$ at $p_{\rm inj}$ and $p_{\rm max}$, respectively, 
while stretching to higher $p_{\rm max}(t)$.
The volume integrated distribution function, $G(p)$, also displays 
a similar behavior as $g_s(p)$.
In the bottom panels of Figure 4, $G(p)/t$ and $G(Z)/t$ are shown,
noting that the kinetic energy passed through the shock front
increases linearly with time. 

In Paper I, based on the DSA simulation results for $t\le10$,
we suggested that the distribution function may become 
self-similar in terms of the momentum similarity variable, $Z$, defined
in \S 1. 
If we define the ``partial pressure function'' as
\begin{equation}
F(Z) \equiv g(Z) { p \over \sqrt{p^2+1}}
\ln\left({p_{\rm max} \over p_{\rm inj}}\right), 
\label{Fz}
\end{equation}
then the CR pressure is given by $P_c \propto \int_0^\infty F(Z)dZ.$
We suggested there that the postshock CR pressure stays
constant because the evolution of $F(Z)$ becomes self-similar.
As can be seen in Figure 4 (b)-(c),
the functions $g_s(Z)$ and $F_s(Z)$ at the subshock seem to 
change very slowly, giving the false impression that $F_s(Z)$ might
be self-similar in terms of the variable $Z$.
However, the constant shape of $F(Z)$ cannot be compatible with
the self-similar evolution of the precursor and shock profile. 
Since $f_s(p)\propto (p/p_{\rm inj})^{-q_s}$ at $Z\sim 0$
and $ f_s(p)\propto (p/p_{\rm max})^{-q_t}$ at $Z\sim 1$  
with constant values of $p_{\rm inj}$,  $q_s$, and $q_t$,
the shape of $F(Z)$ should evolve accordingly in the self-similar stage 
(see Fig. 9 below). 

Figure 5 shows how the evolution of $g_s(p)$ depends on the diffusion 
coefficient and preshock temperature, while
other parameters, $M_0=10$, $\epsilon_B=0.2$, and $\theta=0.1$, are fixed.
The same set of models is shown as in Figure 3.
The shape of $g_s(p)$ is somewhat different among different models,
although it seems to remain similar in time for a given model.
The causes of such differences can be understood as follows.
First of all, the value of $g_s(p_{\rm inj})\approx g_{s,th}(p_{\rm inj})$
depends on the value
of $p_{\rm inj} \propto (u_s/c) \propto M_0 \sqrt{T_0}$. 
Secondly, the numerically realized ``effective'' value of the injection
momentum depends on the diffusion coefficient and grid spacing, 
leading to slightly different injection rates and shock structures. 
Thus the postshock $P_{c,2}$ and the compression
ratios (\ie the shock structure) depend weakly on diffusion coefficient,  
as shown in Figure 1 (e)-(f).
The ensuing CR spectra have slightly different values of $q_s$ and
$q_t$ as shown in Figure 5. 

The spectral slope of the CR spectrum is determined by the mean velocity jump 
that the particles experience across the shock structure.
Here we examine how the precursor velocity profile depends on 
the diffusion model.  
Figure 6 (a) shows the velocity structure $U(\xi)=-u(\xi)$ 
in the precursor ($\xi > 0$) for five different diffusion models, where $u(\xi)$ is
defined as shown in Figure 2. 
We use the velocity data in the finest-level grid
as well as in the base grid.
The velocity profiles are quite similar in all the models
except that the model with $\kappa\propto p^{1/2}$ shows a slightly
different pattern at small scales ($\log \xi < -5$).

Since the particles with momentum $p$ feel on average 
the velocity jump over the corresponding diffusion length,
we can find the velocity $U(\xi_p)$ at the distance from the shock that satisfies 
$ x=l_d(p) = \xi_p \cdot (u_{s,i}t)$.
Using equation (\ref{p1a}), we find 
then $\xi_p = f_c (u_s/u_{s,i}) (p/p_{\rm max})^{\alpha}$. 
Then the particles with the same ratio of $p/p_{\rm max}$ diffuse over the
same similarity scale, $\xi_p$, and feel the same velocity
jump, $U(\xi_p)+u_w(\xi_p)-U_2$ across the shock.
Thus the spectral slope can be estimated from the velocity profile as
\citep[\eg][]{be99}
\begin{equation}
q_u (p) = q_u(\xi_p) = {3(U+u_w) \over {U+u_w-U_2}} 
+ {d\ln (U+u_w-U_2) \over {d \ln p}}.
\label{qu}
\end{equation}
Figure 6 (b) shows the spectral slope, $q_u$, 
which is calculated from numerical results of $U+u_w$ for different models. 
These curves compare to the $q(p)$ curves in Figure 5.

The numerical convergence issue should be discussed here.
The base grid had a spatial resolution
$\Delta x_0 = 2\times 10^{-3}$ in the code units.
The small region around the subshock was refined with a number of levels 
increasing to eight, giving there a spatial resolution
$\Delta x_8 = 7.8\times 10^{-6}$. This structure was sufficient to produce
dynamically converged solutions as discussed in Paper I.
The diffusion length near $p_{\rm inj}\approx 10^{-2}$ is, for instance, 
$l_d(p_{\rm inj}) \approx \kappa(p_{\rm inj})/u_{s,i} \approx 10^{-8}$
in T6P1d model and $\l_d(p_{\rm inj}) \approx 2\times 10^{-5}$ 
in T6P1/2 model, where all quantities are given in the code units.
So the solution for equation (\ref{diffcon}) is not resolved for the 
lowest energy particles in T6P1d model, while it should be well 
resolved in T6P1/2 model.
Since low energy particles cannot see the flow structure shorter than 
the minimum numerical thickness of the subshock, \ie $\Delta x_8$, 
corresponding to the effective diffusion length of $p\sim10$ for T6P1d model,
all particles below $p<10$ feel the same subshock compression,
independent of their diffusion lengths.
This leads to a more or less constant $q(p)\approx q_s$ for $p<10$.
The models shown in Figures 4 and 5 exhibit this trend except T6P1/2 model
in which the diffusion of the injected particles are well resolved
with $\Delta x_8/l_d(p_{\rm inj})=0.4$.

The momentum integration of $g(x,p)$, \ie the CR pressure, 
is self-similar in the spatial similarity variable $\xi$.
Moreover, the CR distribution at the subshock, $g_s(Z)$, 
and the volume integrated distribution, $G(Z)$, both change very slowly
in time, when they are expressed in terms of $Z$.
So we expect that the distribution function $g$ in the plane of
$(\xi,Z)$ should change only secularly during the self-similar stage,
although, as mentioned before, $g(Z)$ does not evolve self-similarly 
in the $Z$ space (Fig. 7).  
The phase space distribution of $g(\xi,Z)$ shows that 
most of low energy particles ($Z<0.5$) are confined within
$-0.2 \la \xi \la 0.1$, while the highest energy particles ($Z\sim 1$)
diffuse over $-1 \le \xi \le 1$.
Thus far away from the subshock, both downstream and upstream, 
relativistic particles dominate the CR energy spectrum. 

\subsection{Analytic approximation for CR spectrum}

Based on the results of DSA simulations described in the previous
subsections,
we suggest that the CR spectrum at CR shocks with $M_0\ga 10$
in the self-similar stage 
can be approximated by
the sum of two power-law functions with an exponential cutoff as follows:
for $p_{\rm max} \gg 1 \gg p_{\rm inj}$,
\begin{equation}
g_s(p)= \left[ g_0 \cdot \left ({p \over p_{\rm inj}}\right)^{-q_s+4} ~+~
g_1 \cdot  \left({p \over p_{\rm max}}\right)^{-q_t+4} \right]
\exp\left[ -\left({p \over{1.5p_{\rm max}}}\right)^{2\alpha}\right],
\label{ganal}
\end{equation}
where $q_s > 4$ and $q_t < 4$.
The specific functional form of the exponential cutoff was found
by fitting the numerical simulation results (see. Figs. 4-5).
We have shown that, after the precursor has developed fully,
the CR pressure at the subshock approaches a time-asymptotic value,
which leads to the self-similar evolution of the entire shock structure. 
Then the parameters, $p_{\rm inj}$, $q_s$ and $q_t$ as well as
$g_0 \approx g_{s,th}(p_{\rm inj})$, become constant in time. 
Also, the value of $g_1$ seems to stay roughly constant, 
according the simulation results. 
We will show below $g_1$ has to be approximately constant,
if $P_{c,2}$ remains constant during the self-similar stage. 
Then the only time-dependent parameter in equation (\ref{ganal}) is 
$p_{\rm max}(t)$, which  can be estimated from equation (\ref{p1a}).

Now let us examine how $P_{c,2}$ evolves in time 
with the proposed form of $g_s(p)$ as $p_{\rm max}$ increases to large values.
Adopting $\alpha=1$, the contributions due to the low and
high energy components can be calculated as
\begin{eqnarray}
\label{pcr}
P_L \equiv  \int_{p_{\rm inj}}^{p_{\rm max}} g_0
\left({p \over p_{\rm inj}}\right)^{-q_s+4} 
\exp\left[ -\left({p \over{1.5p_{\rm max}}}\right)^2\right]
{p \over \sqrt{p^2+1}} {dp \over p}, \nonumber \\ 
P_H \equiv  \int_{p_{\rm inj}}^{p_{\rm max}} g_1
\left({p \over p_{\rm max}}\right)^{-q_t+4} 
\exp\left[ -\left({p \over{1.5p_{\rm max}}}\right)^2\right]
{p \over \sqrt{p^2+1}} {dp \over p}.
\end{eqnarray}
In Figure 8, we show the values of $P_L/g_0$ and $P_H/g_1$
as a function of $p_{\rm max}$
for several values of $q_s$ and $q_t$ and $p_{\rm inj}= 10^{-2}$. 
In $M_0=10$ shocks the typical values of the compression ratios are 
$\sigma_s \approx 3.1$ and $\sigma_t \approx 5.0$, so $q_s \approx 4.4$ and 
$q_t \approx 3.75$.
The plot shows that both $P_L/g_0$ and $P_H/g_1$ become constant
as $p_{\rm max}$ becomes ultra-relativistic,
if the shock flow is modified so that $\sigma_s \rightarrow 3$  and
$\sigma_t \gg 4$.
This explains why $P_{c,2}$ approaches an asymptotic value 
as $p_{\rm max}$ becomes large, leading to the self-similar
evolution stage, after the subshock weakens to the subshock Mach number,
$M_1 \sim 3-4$
and the total compression becomes greater than 4.
Therefore $g_1$ should stay constant, if $P_{c,2}$ becomes constant in
the self-similar stage. 

The amplitude $g_1$ can be estimated, if, for example, $P_{c,2}$ 
is known from the DSA simulations; \ie 
the CR pressure obtained with the proposed analytic form of $g_s$
should be equal to the value of $P_{c,2}$ from the DSA simulations.
Alternatively, as outlined in the appendix,
empirical scaling relations established from simulations
can connect $P_{c,2}$ through simple physics to basic shock parameters.
Then all the parameters necessary to construct approximations 
to the CR distribution 
function as given in equation (\ref{ganal}) at arbitrary 
time $t$ are known for the self-similar evolution stage.
Since the time-asymptotic, self-similar solution of {\it evolving} CR shocks cannot be
found (semi-)analytically either from the conservation equations or from the
boundary conditions, we have to rely at least in part on numerical simulations to estimate
the parameters $p_{\rm inj}$, $g_0$, $\sigma_s$, $\sigma_t$, 
and $P_{c,2}$ for given shock parameters. 
The analytic fitting forms that can approximate the DSA simulation results
are described in the appendix. 

In Figures 4 and 5, we compare the analytic fitting formula
in equation (\ref{ganal}) with the results of our DSA simulations.
They show good agreement.
These plots also demonstrate that $g_s(p_{\rm max})$, and therefore,
$g_1$, remains constant in the self-similar evolution stage. 
The compression ratios shown in Figure 1 are
$\sigma_s \approx 3.2$ and $\sigma_t \approx 5.0$, so the
power-law indices calculated with these ratios are $q_s=4.36$ and $q_t=3.75$.
But the numerical value of $q= - d \ln f_s/ d \ln p$ near $p_{\rm inj}$ is 4.2,
because the diffusion of low energy particles is not resolved fully.  
The minimum value of  $q= - d \ln f_s/ d \ln p$ near $p_{\rm max}$ is 3.79, 
slightly larger than $q_t$, because of the exponential cutoff.
Just to demonstrate how the proposed form of $g_s(p)$ fits the simulation 
results, we use $q_s=4.2$ and $q_t=3.76$ instead for the curve shown in
Figure 4. 
We note that \citet{be99} suggested the minimum value of $q$ is
$q_{\rm min}=3.5 + (3.5-0.5\sigma_s)/ (2\sigma_t - \sigma_s -1)$.
With our compression ratios, $\sigma_s=3.2$ and $\sigma_t=5.0$, this
gives $q_{\rm min}=3.83$, which is slightly larger than our estimate of 3.79.

Using equations (\ref{p1a}) and (\ref{ganal}),
we can estimate the CR spectrum $g_s$ at arbitrary time in the self-similar stage,
as demonstrated in Figure 9 .
Here the value of $g_1$ is fixed by setting $P_{c,2}=0.30$ at $t =1$
and then the same value of $g_1$ is used for the time $t>10$.
From the curves of cumulative $F_s(<Z)$, we can see that
$P_{c,2}$ stays almost constant with the constant value of $g_1$, 
even though $p_{\rm max}$ increases five orders of magnitude.
In fact, $P_{c,2}/(\rho_0 u_{s,i}^2)$ increases from 
0.30 to 0.32 as $p_{\rm max}$ increases from $10^5$ to $10^{10}$.
For such a long span of time, however, $g_s(Z)$ or
$F_s(Z)$ does not keep the same shape. 
At $t=10^5$, the maximum momentum corresponds 
to $p_{\rm max}\approx 10^{19} ({\rm eV}/c)$ for protons.

One might ask how we can justify the validity of the proposed form
of $g_s$ at $t\gg 1$, while our DSA simulations have been carried
up to $t\sim 10-20$. In the T6P1d model, $p_{\rm max}\sim 10^6$ at $t= 10$.
So, most CRs are already ultra-relativistic, and the CR
spectrum evolves as expected (\ie according to eq. [\ref{ganal}]).
As long as $P_{c,2}$ stays constant, the self-similarity of
the precursor/subshock structure would be preserved even for $t\gg 1$.
The stretching of the $u(x)$ profile in the precursor should influence the slope of
the CR spectrum in a self-consistent way as shown in Figure 6. 
There is no physical reason why such feedback between the precursor 
structure and the CR spectrum cannot be extended to $t \gg 1$,
as long as the assumed CR diffusion model remains valid
and the most energetic particles remain contained within the system.
In realistic shocks, however, the assumption for Bohm diffusion could break down
due to inefficient generation of waves in the precursor.
Moreover, highest energy particles escape from the system,
when their diffusion length becomes larger that the physical extent of
the shock.
The effects of escaping particles will be explored further in the next section.

We have focused here on moderately strong shock evolution with $M_0 \ga 10$, since
it is much more complicated to study nonlinear DSA at weaker shocks with $M_0 < 10$.
Nonrelativistic CRs play a more significant role within
those shocks. For instance, since $P_c$ is not dominated by relativistic CRs,
we need to follow more accurately the diffusion of nonrelativistic 
particles on scales close to the physical subshock thickness.
Consequently, the diffusion model and the numerical grid resolution
become important.
The solutions also depend sensitively on the injection momentum,
especially for shocks with Mach numbers, $M_0 \la 2.5$, where modifications
are small, so the nearly test-particle CR spectrum is largely controlled
by the injection momentum.
Physics of thermal leakage injection, however, is not fully understood yet
and we have only a working numerical model.
Thus we defer discussion of semi-analytic discussion of
evolving weak CR shocks to a separate paper.

\subsection{Steady State Shocks with a fixed $p_{\rm ub}$}

In realistic shocks,
$p_{\rm max}(t)$ may reach an upper momentum boundary, $p_{\rm ub}$,
beyond which CRs escape upstream from the shock 
due to the diffusion length, $l_{\rm max}$, approaching the physical
size of the shocked system,  
or to lack of scattering waves at resonant scales of most energetic particles.
From that time the precursor will cease
to increase in scale and the self-similar evolution makes a transition
into a stationary shock structure, or the one controlled by the overall
dynamics of the situation.
Because the shock energy is lost through particles escaping the system
beyond $p_{\rm ub}$,
the self-similar broadening of the precursor is replaced by
a constant precursor structure in steady state. 

We have calculated additional runs for the T6P1d model
in which an upper momentum boundary condition, \ie
$g(p)=0.0$ for $p\ge p_{\rm ub}$ is enforced.  
In these simulations once $p_{\rm max}(t)$ has reached
the given value of $p_{\rm ub}$, the highest energy particles escape
from the shock, the CR spectrum becomes steady and the precursor stops growing.
Figure 10 shows the results of T61Pd model with $p_{\rm ub}=10^5$ and
without the upper momentum boundary.
The distribution function $g_s(p)$ at the shock as well as 
the precursor and subshock structures all become steady after $t>1$
in the run with $p_{\rm ub}=10^5$.
In the other run without particle escape, 
the precursor continues to broaden and $p_{\rm max}(t)$ increases with time.  
However, the postshock states (\eg~ $\rho_2$ and $P_{c,2}$) in the two runs are 
quite similar and $g_s(p)$ in the steady state limit is almost the same as that
of the run without particle escape at $t \approx 1$, except the exponential
tail above $p_{\rm max}$.
In Figure 8 we showed that $P_{c,2}$ stays constant as $p_{\rm max}(t)$ 
increases with time, if $g_s(p)$ follows the form given in equation 
(\ref{ganal}). This explains why $P_{c,2}$ are very similar 
at different times in the two runs.
Minor differences are slightly lower $P_{c,2}$ and higher $\rho_2$ in the
run with particle escape at $p_{\rm ub}$. 
We note that the compression ratio greater than 4 results mainly from the
combined effect of the precursor compression and the subshock jump,
\ie $\sigma_t = \sigma_p \cdot \sigma_s$, regardless of particle escape.
Energy loss due to escaping particles enhances the compression behind
the shock only slightly in this shock, since the loss rate is not significant.

In Figure 11 (a) and (b) snap shots are shown 
at $t=1$ for the runs with $p_{\rm ub}=10^4$ and $10^5$,
and at $t=10$ for the run with $p_{\rm ub}=10^6$. 
For comparison, we also show the time-dependent solutions
at $t=1$ and $10$ for the run without particle escape, 
since in the evolving shock $p_{\rm max}\approx 10^5$ and $10^6$ at $t=1$ 
and 10, respectively, for the T6P1d model.
(At $t=0.1$, $p_{\rm max}$ would reach roughly to $10^4$, but by that time
dynamical equilibrium has not been achieved and the self-similar evolution 
has not begun yet in the simulations.) 
The precursor structure shown in the profile of $P_c$
reflects the diffusion length of highest momenta,
$l_d(p_{\rm ub})\propto p_{\rm ub}$
or 
$l_d(p_{\rm max})\propto p_{\rm max}(t)$.
Here the CR pressure is plotted against $\xi=x/(u_{s,i} t)$, since the results at
two different times are shown together.
So for example, the precursor width in $\xi$ is the same for 
the run with $p_{\rm ub}=10^5$ at $t=1$ (dashed line) and 
the run with $p_{\rm ub}=10^6$ at $t=10$ (long dashed line).
Compared to these two runs, 
the run without particle escape at $t=1$ and 10 (solid lines)
have a wider precursor due to the particles in the exponential
tail above $p_{\rm max}(t)$.  
In Figure 11 (c) and (d) we demonstrate that the evolution of 
the shock structure is quite similar and the shock approaches similar asymptotic 
states for all the runs, almost independent of $p_{\rm ub}$ or $p_{\rm max}(t)$, 
which is consistent with Figure 8.
The asymptotic value of $P_{c,2}$ is slightly lower and the precursor width
is smaller in the runs with smaller $p_{\rm ub}$, as expected.
Otherwise, the steady solutions with different $p_{\rm ub}$ are 
approximately the same as the time-dependent solutions at the time $t$ when 
$p_{\rm max}(t)$ equals to $p_{\rm ub}$. 
Thus the proposed form of $g_s(p)$ can be applied to 
steady state shocks with an upper momentum boundary $p_{\rm ub}=
p_{\rm max}$ as well, ignoring the exponential tail above $p_{\rm max}$.
Even in the case where the shock structure is significantly affected
by the energy loss due to escaping particles, equation (\ref{ganal})
can provide the steady state solution for $g_s(p)$,
if the shock structures ($\sigma_s$, $\sigma_t$ and postshock
states) are known.

\section{Summary}

We have studied the time-dependent evolution of the CR spectrum at CR
modified shocks in plane-parallel geometry, 
in which particles are accelerated to ever higher energies;
that is, the maximum momentum $p_{\rm max}$ is not prefixed.
We adopted Bohm diffusion as well as the diffusion with the power-law momentum
dependence of $\kappa(p) \propto p^{\alpha}$ with $0.5\le \alpha\le 1$. 
Thermal leakage injection of suprathermal particles into the CR population 
at the subshock and finite Alfv\'en wave transport are included.
Simulation parameters target nonrelativistic shocks with $M_0\ga 10$ 
in warm photoionized and hot shock-heated astrophysical environments
with magnetic field strengths somewhat below equipartition
with the thermal plasma.

Unlike gasdynamic shocks, the time-asymptotic dynamical state of the 
evolving CR modified
shocks under consideration here cannot be found analytically either from the
conservation equations or from the boundary conditions.
So we rely on the kinetic simulations of diffusive shock acceleration
to find the time-asymptotic state in the self-similar evolution stage.
The general characteristics of the evolution of shock structure and
particle spectrum can be summarized as follows:

1) The width of the precursor, $H$, scales with the diffusion length of the
most energetic particles and for diffusion that scales
as $\kappa= \kappa^* (\rho_0/\rho)^{\nu} p^{\alpha} $, increases linearly with time, \ie
$ H \approx l_{\rm max} \approx 0.1 u_s t$, independent of 
the magnitude ($\kappa^*$)and the value of $\alpha$.

2) If the acceleration time scale to reach relativistic energies from
injection is much shorter than the dynamical time scale of the shock system
(\ie $\kappa^* \ll 0.1 u_s R$, where $R$ is the characteristic size 
of the shock),
the CR pressure at the subshock approaches a constant value as the
$P_c$ at the shock becomes a significant fraction of the
momentum flux through the shock, $\sim \rho_0 u^2_0$.
For typical nonrelativistic shocks associated
with cosmic structure this transition roughly
corresponds to a time when $p_{\rm max}$ becomes ultra-relativistic.
Once this dynamical equilibrium develops, the shock precursor 
compression and the subshock jump are steady,
leading to a self-similar stretching of the precursor with time.
Consequently, the subshock compression ratio, $\sigma_s$, the total compression
ratio, $\sigma_t$, as well as the postshock gas and CR pressures, 
$P_{g,2}$ and $P_{c,2}$, remain constant during the self-similar stage of the shock.  

3) The lowest energy particles diffuse on a scale $l_{\rm min} = \kappa(p_{\rm inj})/u_s$
and, so, experience only the compression across the subshock.
Thus, near the injection momentum, $p_{\rm inj}$, the CR distribution function
is given by $f(p) \approx f_{s,th}(p_{\rm inj})(p/p_{\rm inj})^{-q_s}$ where
$f_{s,th}$ is the thermal Maxwellian distribution of the postshock gas and
$q_s=3\sigma_s/(\sigma_s-1)$. 
The amplitude $f_{th}(p_{\rm inj})$ is determined by
the thermal leakage injection physics, since that establishes $p_{inj}$. 

4) The most energetic particles diffuse on a scale
$l_{\rm max} = \kappa(p_{\rm max})/u_s$
and, so, experience the total compression across the entire shock structure.
Consequently, near $p_{\rm max}$, $f(p)$ flattens to $(p/p_{\rm max})^{-q_t}$, where  
$q_t=3\sigma_t/(\sigma_t-1)$.
For $p>p_{\rm max}$, $f(p)$ is suppressed by an exponential cutoff.

Considering these facts, we proposed that the CR spectrum at the subshock
for arbitrary time $t$ after self-similar evolution begins can be 
described approximately by the following simple analytic formula:
\begin{equation}
f_s(p,t )= \left[ f_0 \cdot \left({p \over p_{\rm inj}}\right)^{-q_s} ~+~
f_1 \cdot \left({p \over p_{\rm max}(t)}\right)^{-q_t} \right] 
\exp \left[ -\left({p \over{1.5p_{\rm max}(t)}}\right)^{2\alpha}\right],
\label{fanal}
\end{equation}
where $f_0=f_{s,th}(p_{\rm inj})$ and $p_{\rm max} \propto
(u_s^2 t/\kappa^*)^{1/\alpha}$ is given in equation (\ref{p1a}).
The parameters, $p_{\rm inj}$, $q_s$ and $q_t$ can be
estimated from the shock structure in the self-similar stage
using DSA simulations results as outlined in the appendix.
The amplitude, $f_1$, has to satisfy the relation 
$g_s(p_{\rm max})=f_s(p_{\rm max}) p_{\rm max}^4 \approx $ constant
in order for the postshock $P_c$ to remain steady.
So, the momentum distribution function $g(p)$ is shifted to higher
$p_{\rm max}$ in time, while keeping the amplitude at $p_{\rm max}$ constant
in the self-similar stage.
Hence $p_{\rm max}$ is the only time-dependent parameter in equation
(\ref{fanal}).

In a realistic shock geometry, however,
CRs may escape upstream from the shock due to largest diffusion length 
approaching the physical size of the shocked system, 
or due to lack of scattering waves at resonant scales of most energetic particles.  
Once $p_{\rm max}$ approaches some upper momentum boundary at $p_{\rm up}$, 
the shock structure and the CR spectrum develop steady states that are approximately
the same as the evolving forms with $p_{\rm max} = p_{\rm up}$,
except that some differences in the shock structure due to energy loss from
escaping particles. 
Otherwise, the shock structure parameters and the approximate analytic form for the CR spectrum 
in the self-similar stage are consistent with previously proposed 
analytic and semi-analytic steady state solutions \citep[\eg][]{be99,ab05}.

Finally, we note that the evolution of the CR spectrum is secular 
in terms of the variable, $Z= \ln(p/p_{\rm inj})/\ln(p_{\rm max}/p_{\rm inj})$,
which alluded wrongfully the self-similar evolution of the partial pressure
function $F_s(Z)$ in Paper I. 
In fact there is no similarity relation between $p$ and $t$.

\acknowledgements
HK was supported by the Korea Research Foundation Grant
funded by the Korean Government (MOEHRD) (R04-2006-000-100590).
DR was supported by the Korea Research Foundation Grant
funded by the Korean Government (MOEHRD) (KRF-2007-341-C00020).
TWJ is supported at the University of Minnesota
by NASA grant NNG05GF57G, NSF grant Ast-0607674
and by the Minnesota Supercomputing Institute.

\appendix
\section{Analytic approximations for dynamical states}

As we noted in the Introduction, there are several analytic 
and semi-analytic treatments of strong, steady-state CR modified shocks. 
The full time-asymptotic state of evolving CR modified shocks 
can be obtained only through numerical simulations of nonlinear DSA. 
However, such simulations
show strong similarities between steady-state and asymptotic, evolving shocks.
Here we outline some of those basic dynamical relations as they 
can be estimated analytically
and empirically from our simulations, as reported in this paper 
and previously in Paper I.

A key to this comparison is the fact that the time scale for evolution of the shock
precursor is the acceleration time scale to reach $p_{\rm max}$, 
$t_{\rm acc} \sim 10 (l_{\rm max}/u_s)$ (see eq. [\ref{tacc}]),
which is characteristically an order of magnitude greater 
than the time scale for a fluid
element to pass through the precursor, $t_{\rm dyn} \sim l_{\rm max}/u_s$. 
Then, in following a fluid element through the precursor, 
one can neglect terms $\partial/\partial t$
compared to terms $u\partial /\partial x$ in evaluating 
the Lagrangian time variation, $d /dt$.
For example, equation (\ref{econ}), which can be expressed as
\begin{equation}
\frac{d}{dt}\left(\frac{P_g}{\rho^{5/3}}\right) = 
\frac{2}{3}\frac{W}{\rho^{5/3}},
\end{equation}
assuming $\gamma_g = 5/3$, then gives for an evolving precursor 
\begin{equation}
P_{g,1} \approx \left( P_{g,0} + \frac{2}{5}\rho_0 u^2_0 I\right)
\sigma_p^{5/3},
\label{pp}
\end{equation}
where $\sigma_p = \rho_1/\rho_0$ is the precursor compression factor.
The quantity
\begin{equation}
I = \frac{5}{3 u^3_0 \rho^{1/3}_0} \int \frac{|W|}{\rho^{2/3}} dx
\label{wavei}
\end{equation}
was introduced in Paper I, and measures  entropy added by 
Alfv\'en wave dissipation
while the fluid element crosses the precursor, 
normalized by $u^2_0\rho_0/\rho^{5/3}_0$. 
Since equation (\ref{pp}) applies to an evolving shock, the subscripts `0' and
`1' refer to states of a given fluid element as it enters the precursor 
and as it
reaches the subshock. The approximation comes from neglecting 
explicit time variations
in $|W|$ and $\rho$ in evaluating $I$. Equation (\ref{pp}) is 
exact for a steady state shock.
In the absence of Alfv\'en wave dissipation, this equation simply 
states the properties
of adiabatic compression through the precursor, which obviously does 
not depend on the precursor being steady state.

Along similar lines, momentum conservation of a fluid element passing through
the (slowly) evolving precursor gives
\begin{equation}
P_{c,1} + P_{g,1} \approx P_{g,0} + \rho_0 u^2_0 
\left(1 - \frac{1}{\sigma_p}\right),
\label{pcomp}
\end{equation}
which can be combined with equation (\ref{pp}) to 
produce a simple estimate for the
CR pressure at the subshock,
\begin{equation}
P_{c,1} = P_{c,2} \approx \rho_0 u^2_0 \left[ 1 - \frac{1}{\sigma_p} - 
\frac{3}{5}\frac{\sigma^{5/3}_p - 1}{M^2_0} - 
\frac{2}{5}I\sigma^{5/3}_p\right].
\label{pcest}
\end{equation}
By substituting equation (\ref{pcest}) into equation (\ref{wavei})
along with equation (4), one can obtain 
\begin{equation}
I \approx \frac{5}{3} \frac{v_{A,0}}{u_0} \frac{P_{c,1}}{\rho_0 u_0^2},
\label{wavei2}
\end{equation}
where, once again, the approximation reflects neglect of
explicit time variation in the shock structure during passage of
a fluid element through the shock.
Substituting this back into equation (\ref{pcest}) we obtain
\begin{equation}
\frac{P_{c,2}}{\rho_0 u_0^2}\approx \left[{ 1 - \frac{1}{\sigma_p}-
\frac{3}{5}\frac{\sigma^{5/3}_p - 1}{M^2_0} } \right]
\left[ 1+ \frac{2}{3} \frac{v_{A,0}}{u_0} \sigma_p^{5/3} \right] ^{-1}.
\label{pcest2}
\end{equation}
Given $P_{c,1}=P_{c,2}$ from equation (\ref{pcest2}) and using equation 
(\ref{pcomp})
it is straightforward to determine, as well, $P_{g,1}$.

Although we can estimate approximately the postshock pressures, $P_{g,2}$ and $P_{c,2}$,
for a given value of precursor compression, we must rely on numerical simulations to obtain 
the value of $\sigma_p$ for different model parameters.
In the remainder of this appendix we present some practical expressions for the shock 
dynamical properties obtained in our DSA simulations using a wide range of Mach numbers 
for the thermal injection parameter
$\epsilon_B = 0.2$, the Alfv\'en wave transport parameter, 
$\theta = 0.1$ and the
diffusion coefficient, $\kappa = \kappa^* p (\rho/\rho_0)$.
In Figure 11 the time-asymptotic values of postshock CR pressure, 
gas pressure and compression ratios are plotted against the initial
shock Mach number ($ M_0 \ge 1.5$). 

For $M_0 \le 2.5$, the CR modification is negligible, so the postshock 
gas pressure and the shock compression ratios $\sigma_t=\sigma_s$ are 
given by the usual Rankine-Hondo relation for pure gasdynamic shocks. 

For $M_0 > 2.5$, the numerical results for the postshock gas pressure
can be fitted by 
\begin{equation}
{P_{g,2} \over {\rho_0 u_{s,i}^2} } \approx 0.4
\left({M_0\over10}\right)^{-0.4}
\label{pganal}
\end{equation}
The time-asymptotic density compression ratios can be approximated as follows: 
\begin{eqnarray}
\sigma_s  \approx 3.2 \left({M_0\over10}\right)^{0.17}
~~~{\rm for}~ 2.5\le M_0 \le 10,\\
\sigma_s  \approx 3.2 \left({M_0\over10}\right)^{0.04}
~~~{\rm for}~  M_0 > 10, \nonumber
\label{sigmas}
\end{eqnarray}

\begin{eqnarray}
\sigma_t  \approx 5.0 \left({M_0\over10}\right)^{0.42}
~~~{\rm for}~ 2.5\le M_0 \le 10,\\
\sigma_t  \approx 5.0 \left({M_0\over10}\right)^{0.32}
~~~{\rm for}~ M_0 > 10. \nonumber
\label{sigmat}
\end{eqnarray}
We note that the subshock compression depends only weakly on $M_0$,
while the total compression increases approximately as $M_0^{1/3}$.
Even for strong shocks with $M_0$ up to 100, the total compression 
ratio is less than 10, 
because the propagation and dissipation of Alfv\'en waves upstream reduces the
CR acceleration and the precursor compression.

The postshock CR pressure can be fit empirically as follows:
\begin{eqnarray}
{P_{c,2} \over {\rho_0 u_{s,i}^2} } \approx 2.34\times 10^{-2} (M_0-1)^3
~~~{\rm for}~ 1.5<M_0<2.5, \nonumber \\
{P_{c,2} \over {\rho_0 u_{s,i}^2} } \approx {0.58(M_0-1)^4 \over M_0^4}
- {2.14(M_0-1)^3 \over M_0^4} +  {13.7(M_0-1)^2 \over M_0^4} \\
- {27.0(M_0-1) \over M_0^4} + {15.0 \over M_0^4}
~~~{\rm for}~ 2.5\le M_0 \le 100, \nonumber\\
{P_{c,2} \over {\rho_0 u_{s,i}^2} } \approx 0.55 
~~~{\rm for}~ M_0> 100. \nonumber 
\label{pcanal}
\end{eqnarray}
These fits are plotted in solid lines in Figure 11.
Since $\sigma_p = \sigma_t/\sigma_s$, equations (A9) and (A10) 
can be used along with equation (\ref{pcest2}) to estimate $P_{c,2}$
(dotted line in Fig. 11). 

In \citet{kjg02} we showed that the effective injection momentum
is $p_{\rm inj}/p_{th} \approx 2.5$ for $M_0 \ga 10$
for the injection parameter $\epsilon_B=0.2$, 
where $p_{th}= 2 \sqrt{kT_2/m_pc^2}$ 
and $T_2 = (P_{g,2}/\rho_2) (m_p/k)$ is the postshock gas temperature.
Then the thermal distribution at the injection momentum, $g_{s,th}(p_{\rm inj})$,
can be calculated from the Maxwell distribution,
since the postshock gas states, $T_2$ and $\rho_2$, are known.

\clearpage

\begin{deluxetable} {cccc}
\tablecaption{Preshock Temperature and Diffusion Coefficient
in Numerical Models}
\tablehead{
\colhead {Model Name} & \colhead{$T_0$} & \colhead{$\kappa/\hat \kappa$} 
& Description for Diffusion Coefficient}
\startdata
T6P1d & $10^6$K  & $10^{-6}p (\rho_0/\rho)$
& power-law diffusion with $\rho^{-1}$ dependence\\
T6P3/4d &  $10^6$K  & $10^{-6}p^{3/4} (\rho_0/\rho)$ 
& power-law diffusion with $\rho^{-1}$ dependence\\
T6P1 &  $10^6$K  & $10^{-6}p $
& power-law diffusion\\
T6P1/2 & $10^6$K  & $1.78\times10^{-4}p^{1/2}$
& power-law diffusion\\
T4P1d & $10^4$K  & $10^{-5}p (\rho_0/\rho)$
& power-law diffusion with $\rho^{-1}$ dependence\\
T6Bd &  $10^6$K  & $10^{-2}p^2/\sqrt{p^2+1}(\rho_0/\rho)$
& Bohm diffusion with $\rho^{-1}$ dependence\\
T4Bd &  $10^4$K  & $10^{-2}p^2/\sqrt{p^2+1}(\rho_0/\rho)$
& Bohm diffusion with $\rho^{-1}$ dependence\\
\enddata
\end{deluxetable}

\clearpage

\begin{figure}
\plotone{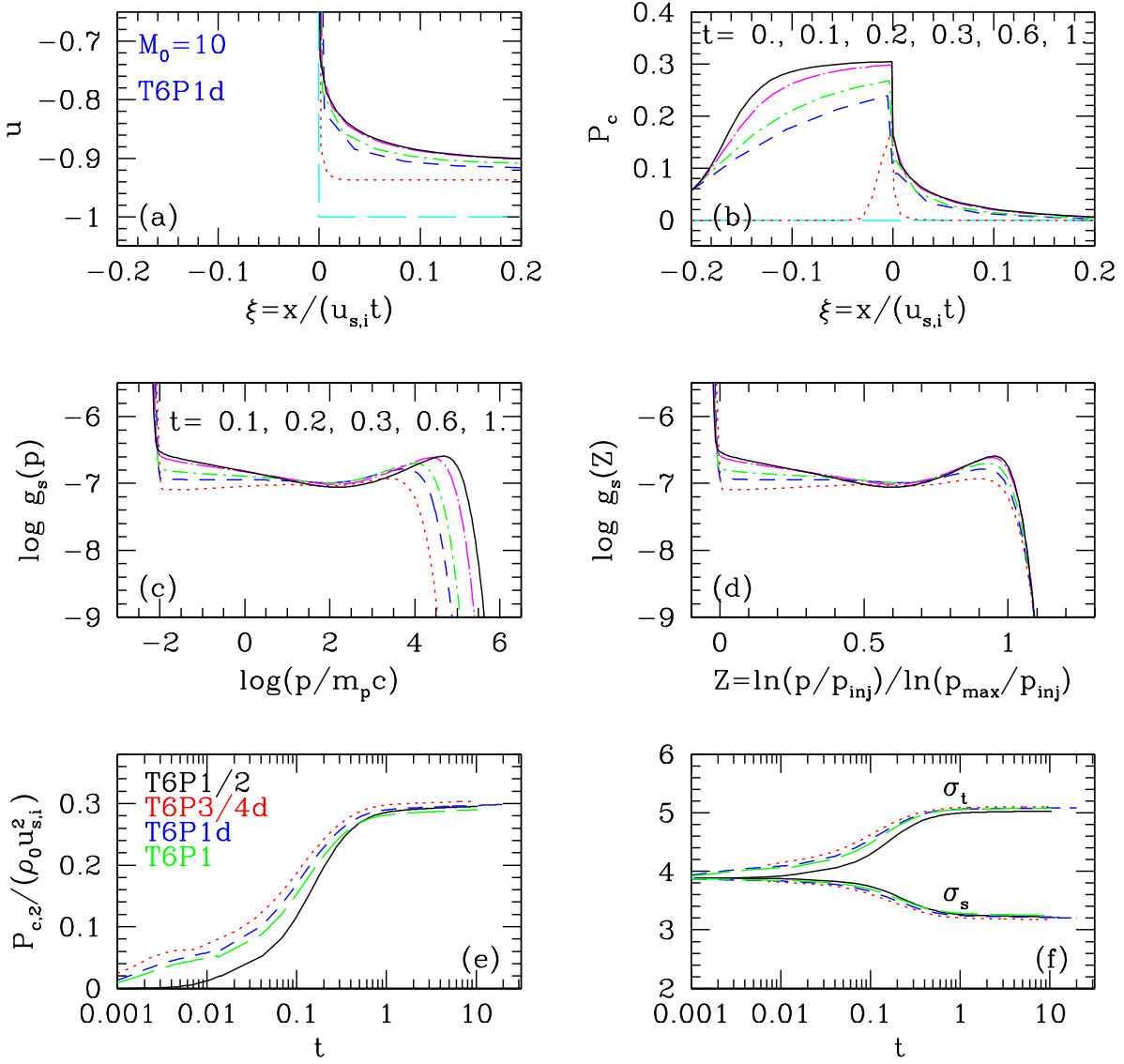}
\figcaption{(a)-(d): Snap shots of $M_0=10$ shock of T6P1d model up to
$t=1$ in terms of a similarity variable $\xi \equiv x/(u_{s,i}t)$.
The flow velocity, CR pressure, and CR distribution function 
at the subshock, $g_s(p)=f_s(p)p^4$, and $g_s(Z)$, are shown 
at $t=$ 0.1 (dotted lines), 0.2 (dashed), 0.3 (dot-dashed), 
0.6 (dot-long dashed), and 1.0 (solid). 
The long dashed lines show the initial shock structure.
(e)-(f): Time evolution of the postshock CR pressure and 
the compression ratios for $M_0=10$ shocks 
with four different models of diffusion coefficient
$\kappa(p)$ (see Table 1).}
\end{figure}

\clearpage

\begin{figure}
\plotone{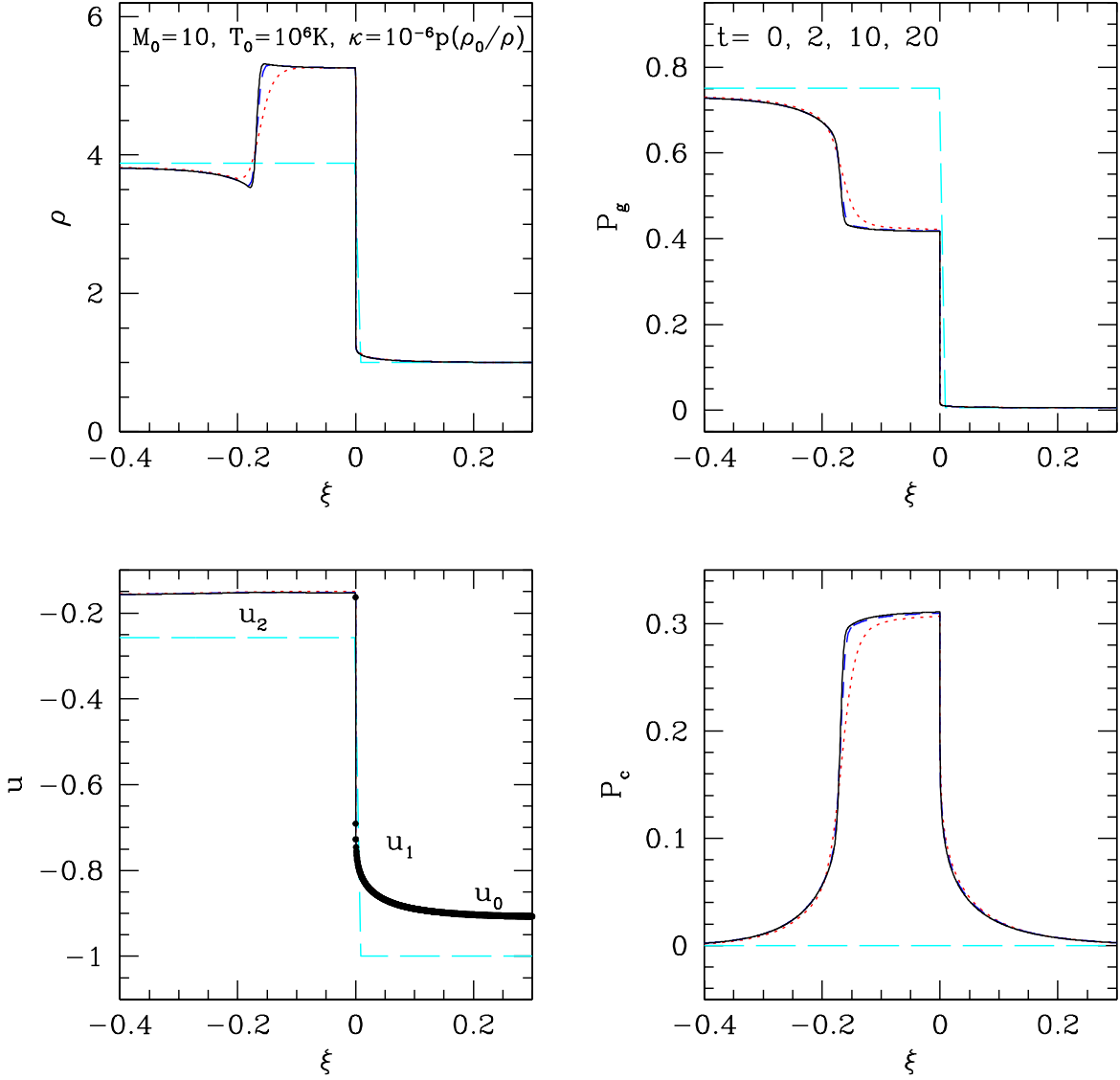}
\figcaption{Self-similar evolution of $M_0=10$ shock of T6P1d model.  
The shock structure is shown at $t=$ 2 (dotted lines),
10 (dashed), 
and 20 (solid) as a function
of the similarity variable $\xi=x/(u_{s,i}t)$ in the shock rest frame.
The long dashed lines show the initial shock structure.}
\end{figure}

\clearpage

\begin{figure}
\plotone{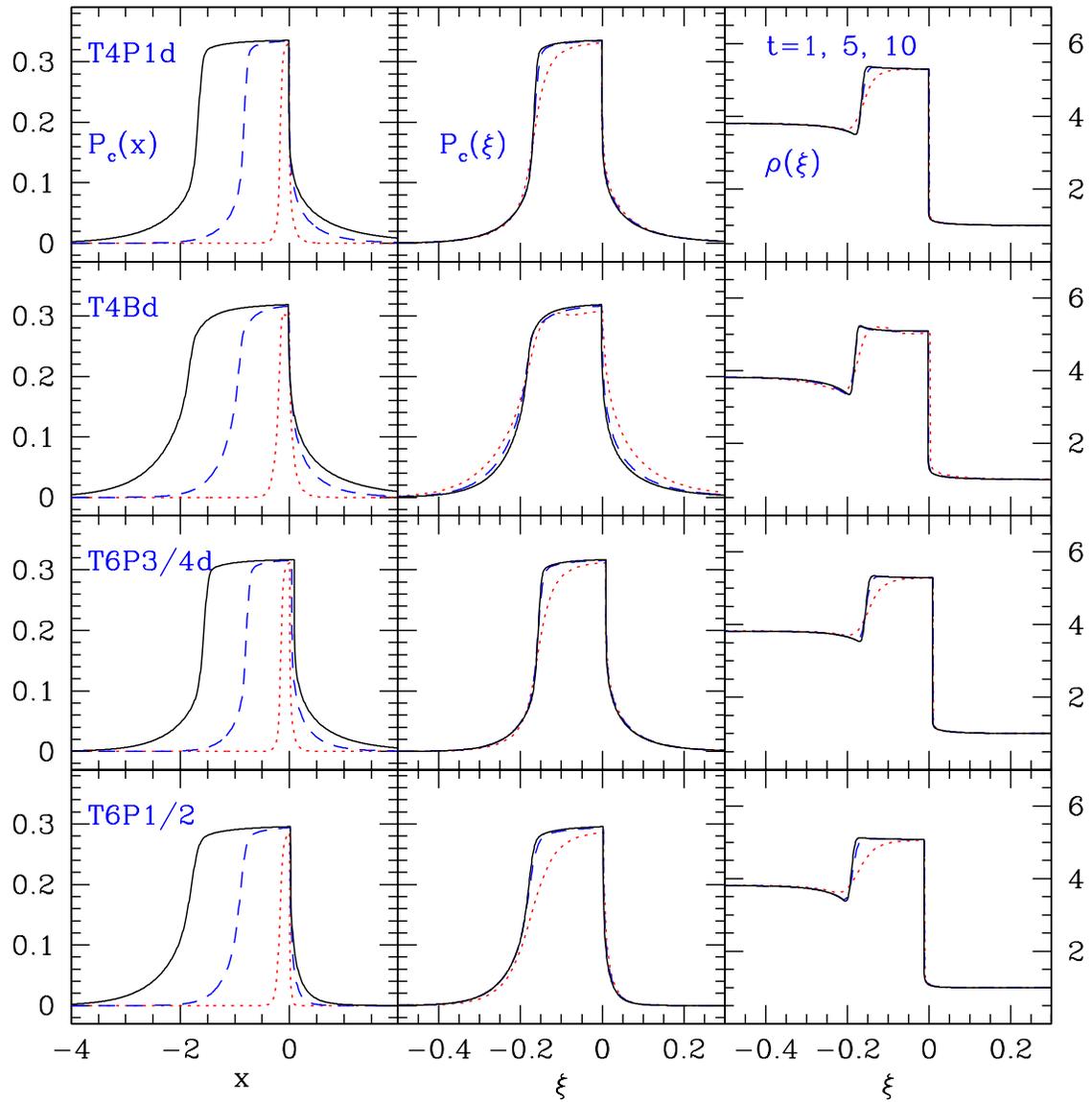}
\figcaption{Self-similar evolution of $M_0=10$ shock of four different
models listed in Table 1, shown at $t=1$ (dotted lines), 5 (dashed), 
and 10 (solid).
The CR pressure is shown as a function of $x$ (left panels) 
and the spatial similarity variable $\xi=x/(u_{s,i}t)$ (middle panels).
The gas density is shown at the right panels.}
\end{figure}

\clearpage

\begin{figure}
\plotone{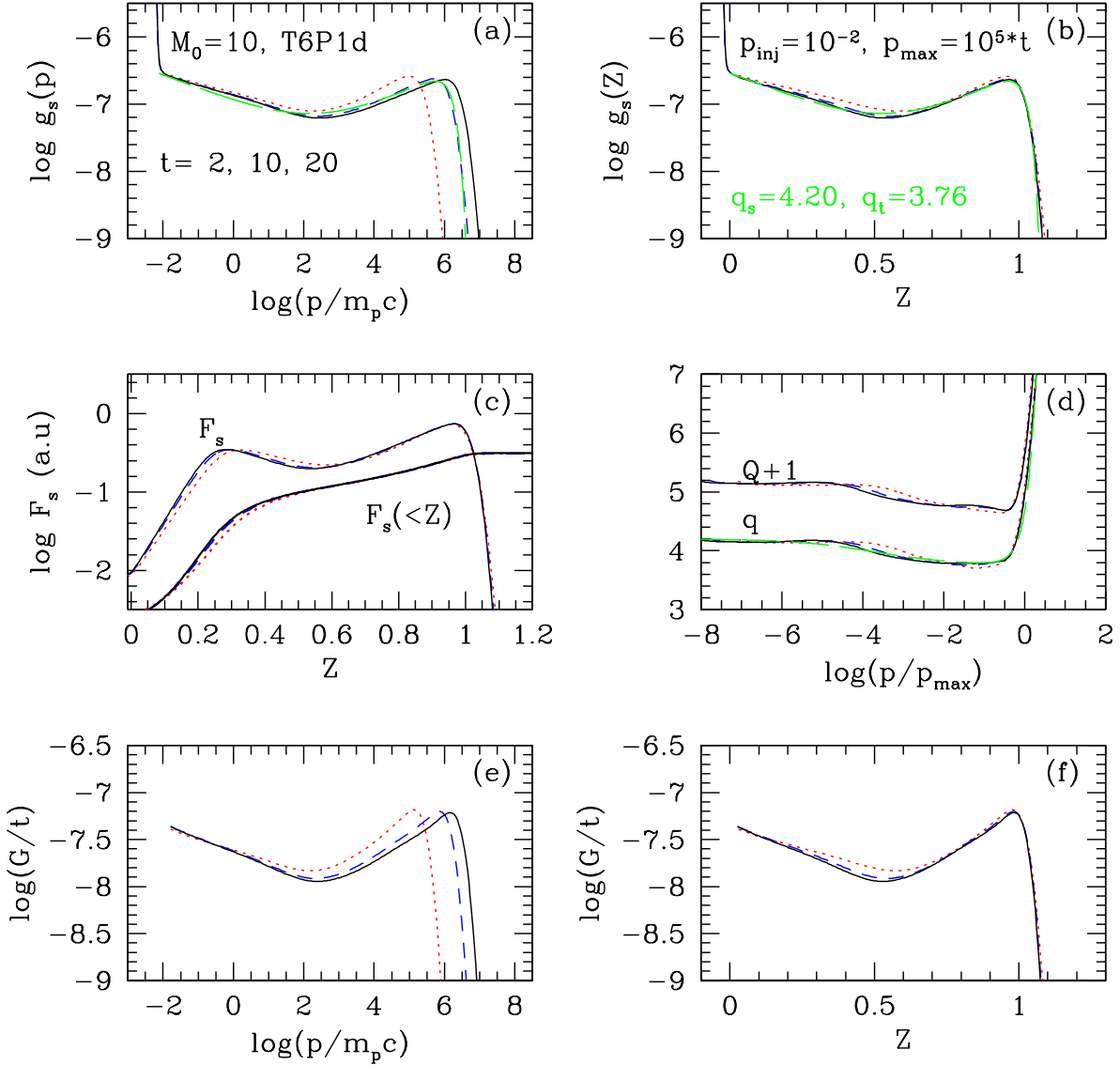}
\figcaption{CR distribution function for $M_0=10$ shock of T6P1d 
model, shown in Fig. 2,
at $t=$  2 (dotted lines), 10 (dashed), and 20 (solid). 
(a)-(b): The distribution function at the subshock, $g_s(p)$ and $g_s(Z)$,
where $Z= \ln(p/p_{\rm inj})/\ln(p_{\rm max}/p_{\rm inj})$.
(c): The partial pressure, $F_s$, defined in equation (\ref{Fz}) 
and its cumulative distribution, $F_s(<z)$. 
(d): The power-law slopes, $q= - d \ln g_s/ d \ln p + 4$ and 
$Q= - d \ln G/ d \ln p + 4$.
(e)-(f): The volume integrated distribution function $G=\int g dx$
plotted against $\log(p)$ or $Z$. 
The long dashed lines in (a), (b), and (d) show the analytic fitting
given in equation (\ref{ganal}) with $q_s=4.20$, 
$q_t=3.76$, $p_{\rm inj}=10^{-2}$, and $p_{\rm max}=10^5 t$ at $t=10$.}
\end{figure}

\clearpage

\begin{figure}
\plotone{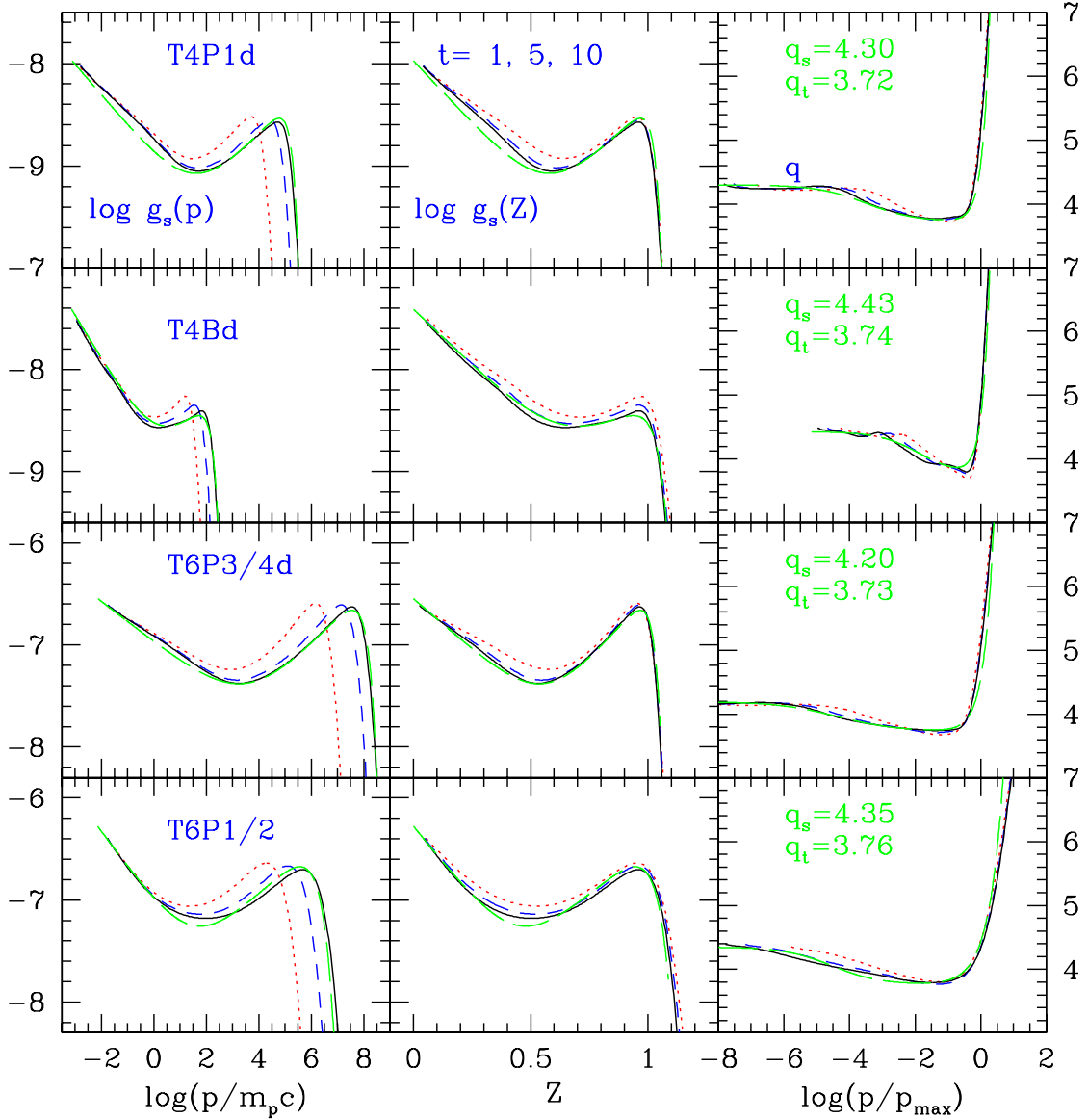}
\figcaption{
CR distribution function at the subshock, $g_s(p)$ and $g_s(Z)$, and 
the power-law slopes, $q= - d \ln g_s/ d \ln p + 4$ are shown
at $t=$  1 (dotted lines), 5 (dashed), and 10 (solid)  
for the four diffusion models shown in Fig. 3.
The long dashed lines show the analytic fitting
given in equation (\ref{ganal}) at $t=10$.
The adopted values of $q_s$ and $q_t$ are given for each model.}
\end{figure}

\clearpage

\begin{figure}
\plotone{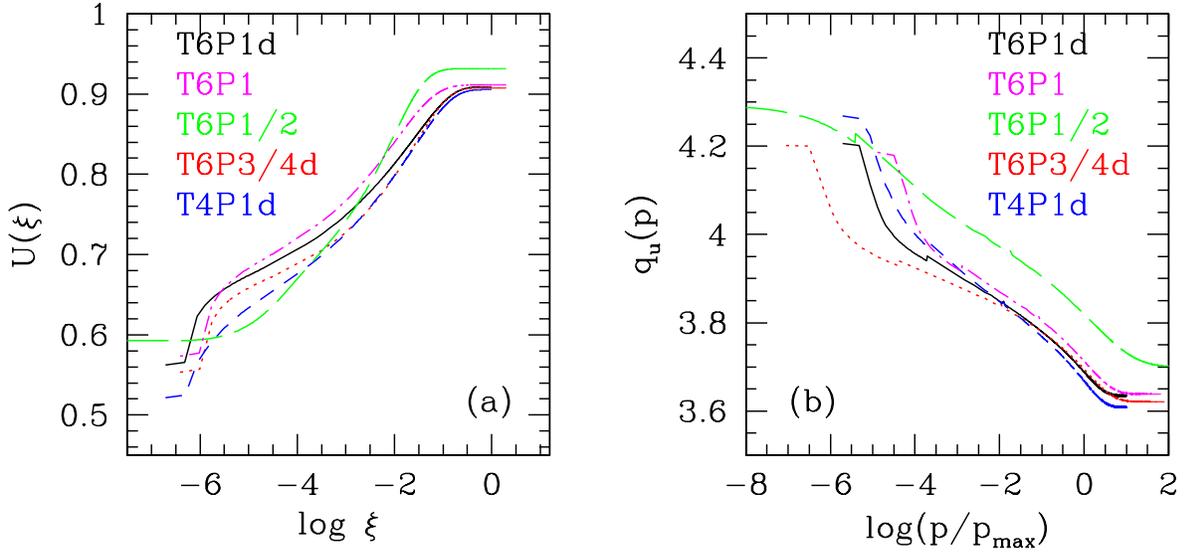}
\vskip -8cm
\figcaption{(a) Velocity profiles in the precursor as a function
the similarity distance from the subshock for five different models with $M_0=10$
listed in Table 1.
(b) The power slope calculated with equation (\ref{qu}) using
the velocity profile shown in (a).}
\end{figure}

\clearpage

\begin{figure}
\plotone{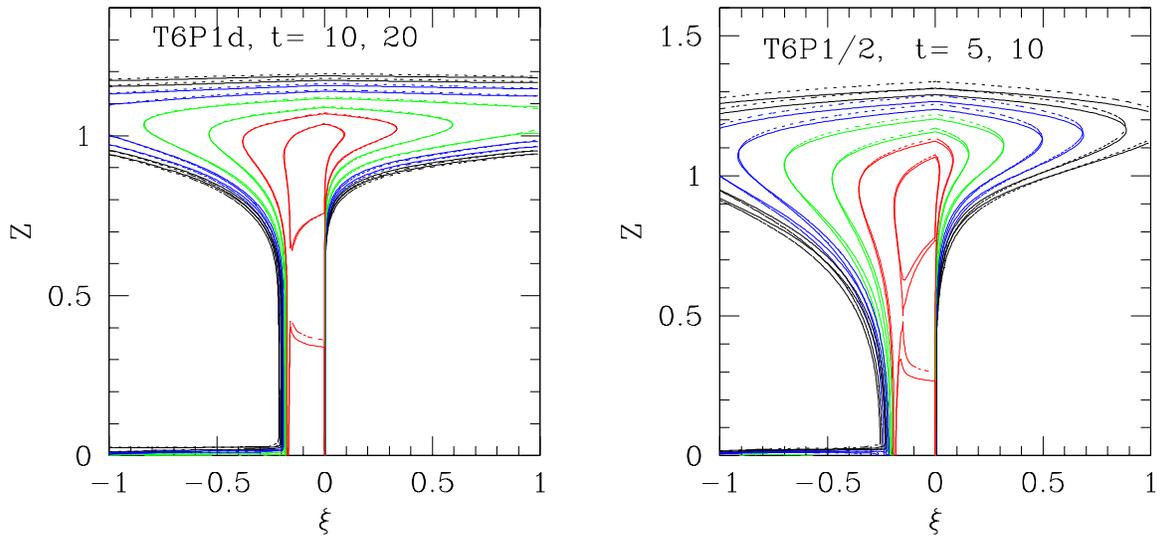}
\vskip -8cm
\figcaption{{\it Left panel:}
Contour plots of $g(\xi,Z)$ at $t=10$ (dotted lines), 20 (solid lines)
for T6P1d model.
{\it Right panel:}
Contour plots of $g(\xi,Z)$ at $t=5$ (dotted lines), 10 (solid lines)
for T6P1/2 model.}
\end{figure}

\clearpage

\begin{figure}
\plotone{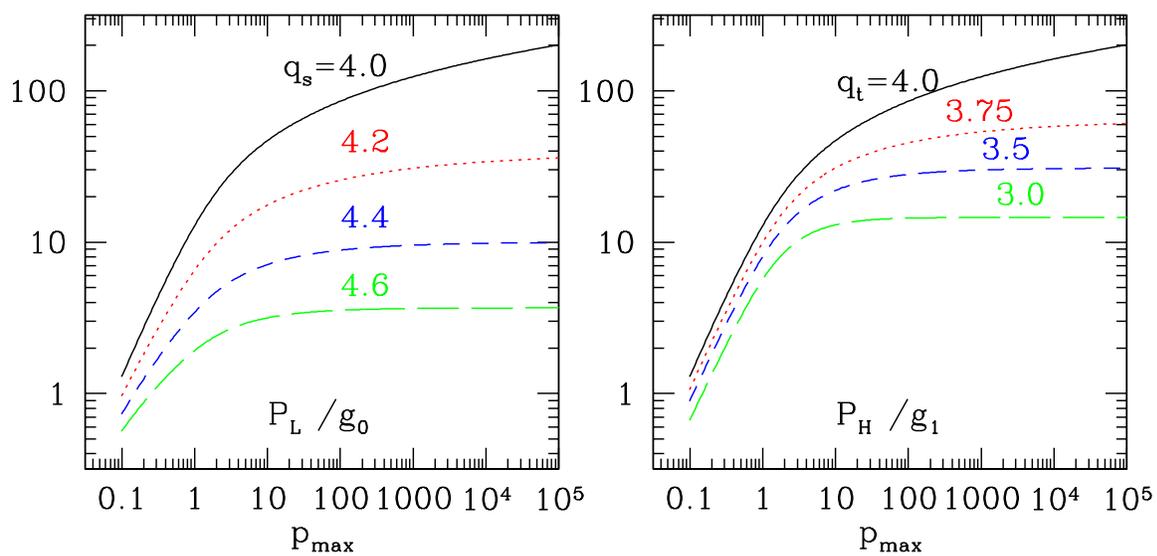}
\vskip -8cm
\figcaption{{\it Left panel:}
$P_L$, defined in equation (\ref{pcr}), for $q_s=4.0-4.6$.
{\it Right panel:}
$P_H$, defined in equation (\ref{pcr}), for $q_t=3.0-4.0$.  
Here the injection momentum is $p_{\rm inj}=10^{-2}$.}
\end{figure}

\clearpage

\begin{figure}
\plotone{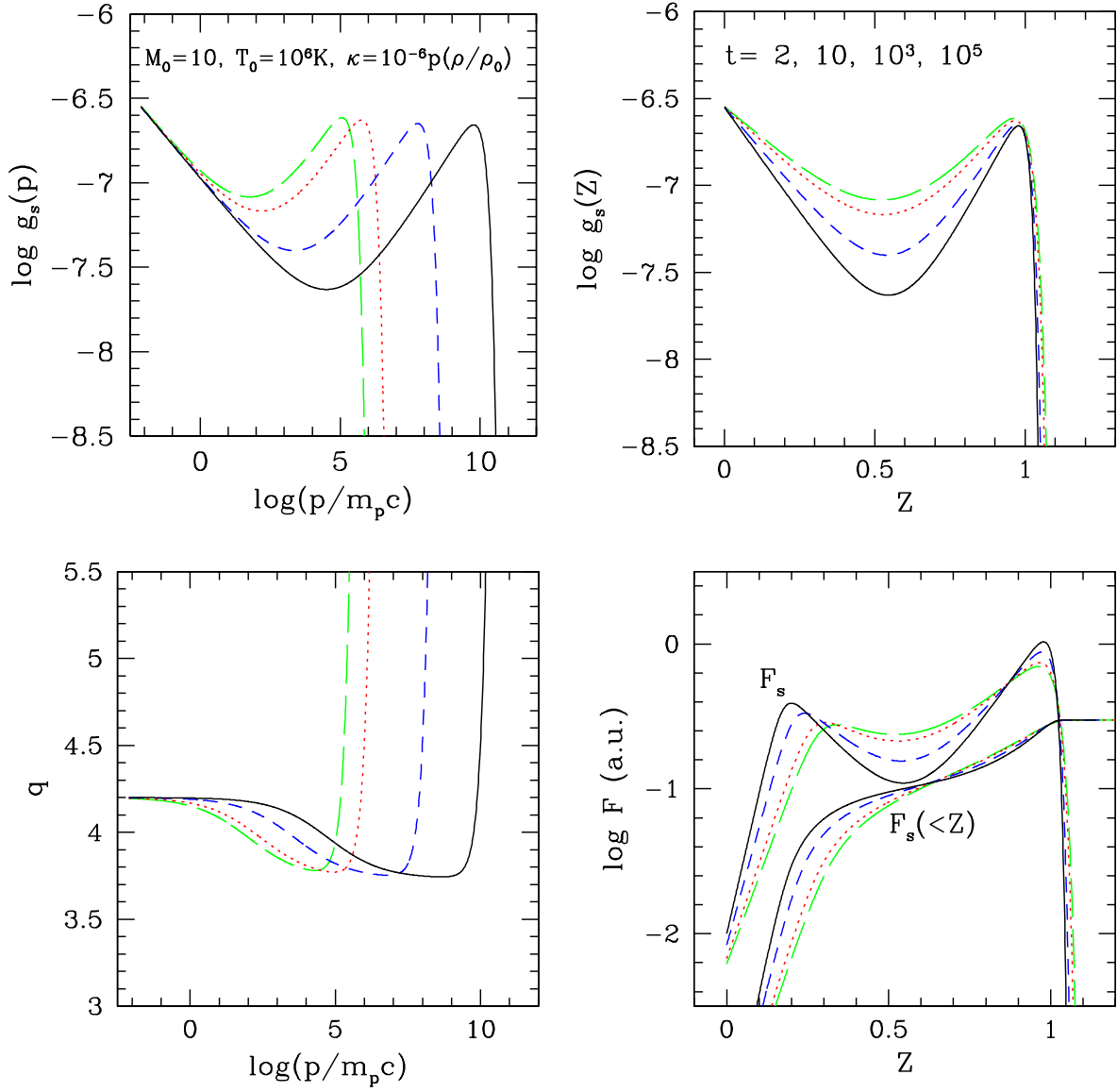}
\figcaption{{\it Upper panels}:
CR distribution at the subshock calculated using the analytic fitting 
formula in equation (\ref{ganal}) with $q_s=4.20$, $q_t=3.76$,
$p_{\rm inj}=10^{-2}$, and $p_{\rm max}=10^5 t$.
{\it Lower panels}: Power-law slope of the fitted $g_s$, \ie
$q= - d \ln g_s/ d \ln p + 4$, and partial pressure, $F_s(Z)$, and
its cumulative distribution, $F_s(<Z)$.}
\end{figure}

\clearpage

\begin{figure}
\plotone{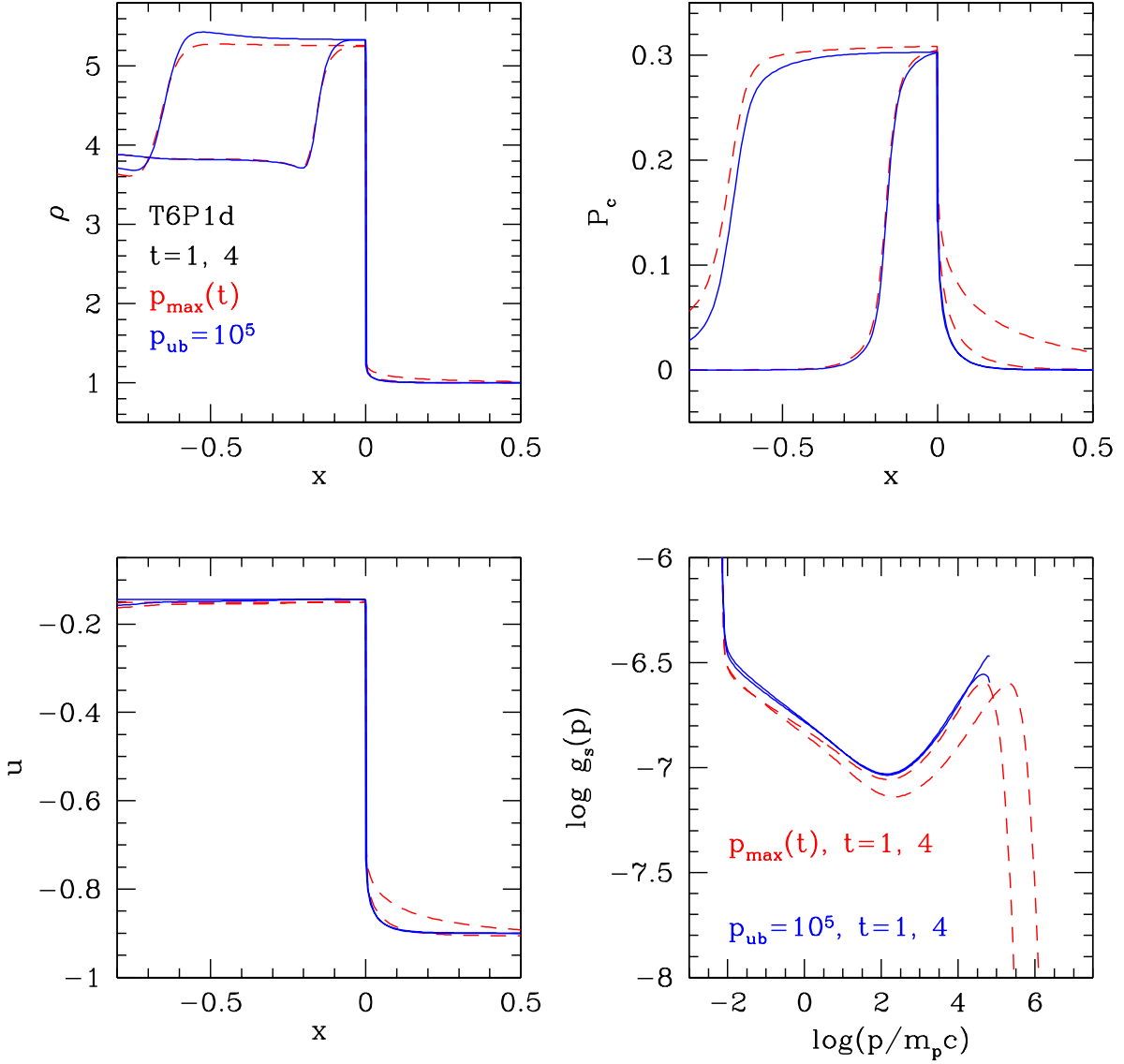}
\figcaption{
Comparison of the run with particle escape at $p_{\rm ub}=10^5$ (solid lines) 
and the run without particle escape (dashed line) for T6P1d model.
The shock structure and the CR distribution are shown at $t=$ 1 and 4.
} 
\end{figure}

\clearpage

\begin{figure}
\plotone{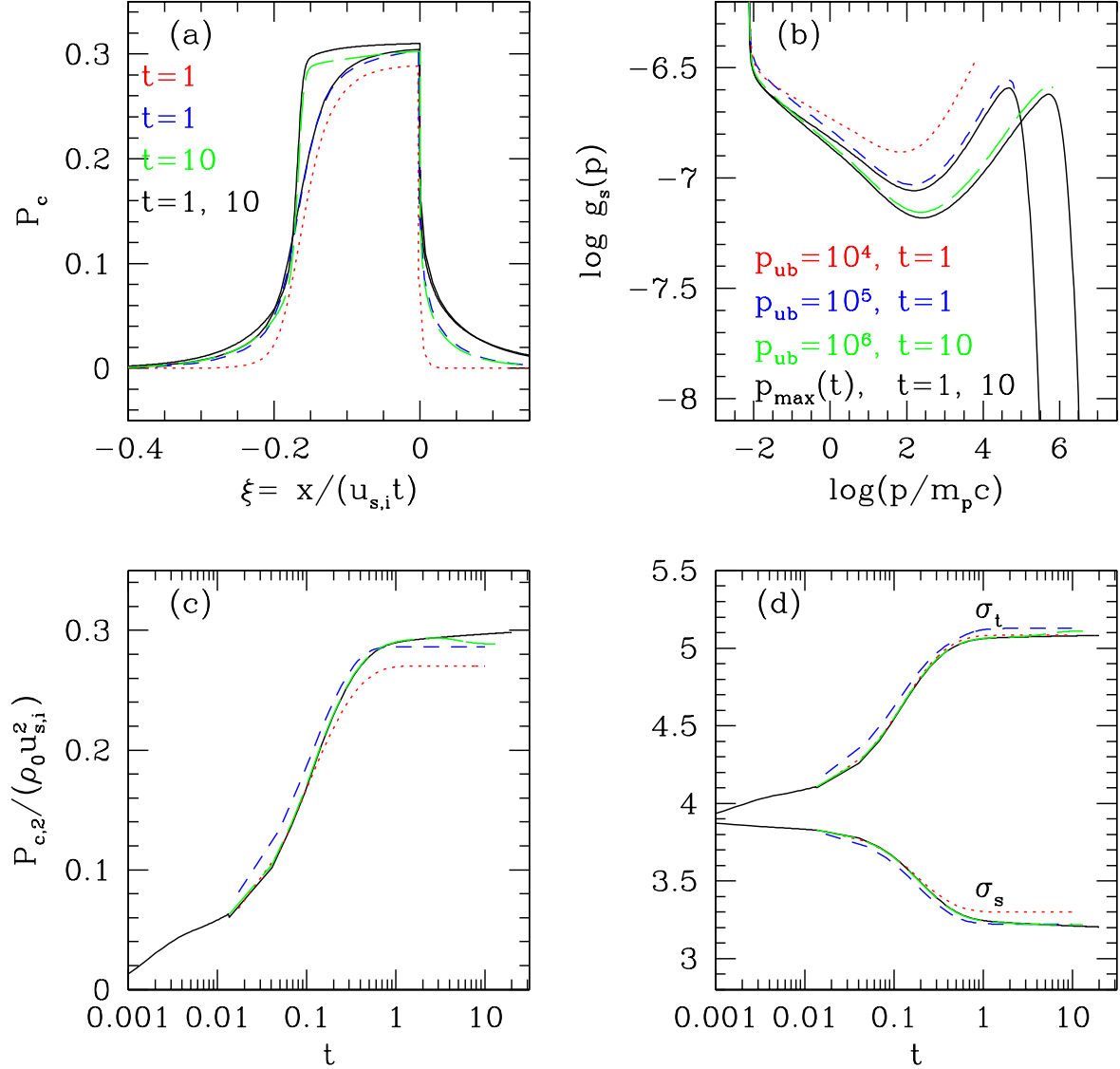}
\figcaption{ 
Comparison of the runs with and without particle escape at $p_{\rm ub}$ 
for T6P1d model.
(a) CR pressure profiles in the three runs 
with $p_{\rm ub}=10^4$ at $t=1$ (dotted line),  
with $p_{\rm ub}=10^5$ at $t=1$ (dashed), and 
with $p_{\rm ub}=10^6$ at $t=10$ (long dashed line). 
The solid lines are for the run
without particle escape at $t=1$ and 10. 
(b) CR spectrum at the subshock. 
(c)-(d): Time evolution of the postshock CR pressure and
the compression ratios.
The same line types are used in all the panels.} 
\end{figure}

\clearpage

\begin{figure}
\plotone{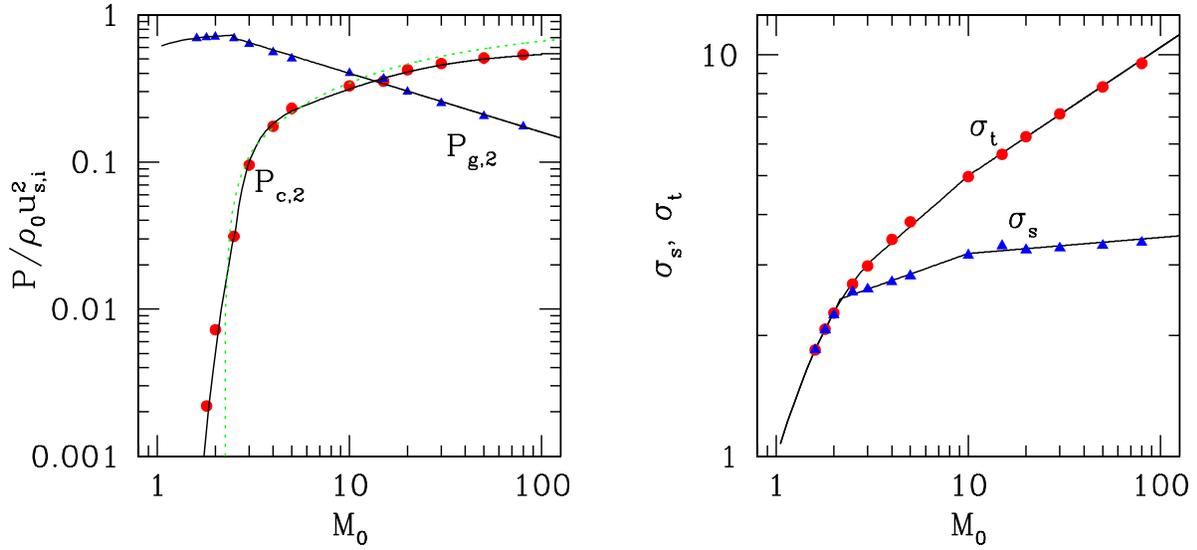}
\vskip -8cm
\figcaption{Time-asymptotic values of postshock gas and CR pressures
in units of initial shock ram pressure (left panel),
subshock compression ratio (triangles, right panel)
and total compression ratio (circles, right panel)
as a function of initial shock Mach number $M_0$ for T6P1d models.
The solid lines show our fitting formulas given in
equations (\ref{pganal})-(\ref{sigmat}).
The dotted line shows the estimate given in equation (\ref{pcest2}),
adopting the numerical values of $\sigma_p=\sigma_t/\sigma_s$ in
equations (\ref{sigmas}) and (\ref{sigmat}). }
\end{figure}

\clearpage

\end{document}